\documentclass[fleqn,usenatbib]{mnras}

\usepackage{newtxtext,newtxmath}

\usepackage[T1]{fontenc}

\DeclareRobustCommand{\VAN}[3]{#2}
\let\VANthebibliography\thebibliography
\def\thebibliography{\DeclareRobustCommand{\VAN}[3]{##3}\VANthebibliography}

\usepackage{graphicx}	
\usepackage{amsmath}	
\usepackage{color}
\usepackage{pdflscape}
\usepackage{subfig}

\def\gtsima{$\; \buildrel > \over \sim \;$}
\def\ltsima{$\; \buildrel < \over \sim \;$}
\def\gsim{\lower.5ex\hbox{\gtsima}}
\def\lsim{\lower.5ex\hbox{\ltsima}}
\def\kms{km s$^{-1}$}

\def\zem{z_{\rm em}}

\def\Lya{Ly$\alpha$}
\def\Lyb{Ly$\beta$}

\newcommand{\CIV}{\mbox{C\,{\sc iv}}}

\newcommand{\CIII}{\mbox{C\,{\sc iii}}}
\newcommand{\CII}{\mbox{C\,{\sc ii}}}
\newcommand{\NV}{\mbox{N\,{\sc v}}}

\newcommand{\OI}{\mbox{O\,{\sc i}}}

\newcommand{\MgII}{\mbox{Mg\,{\sc ii}}}

\newcommand{\SiIV}{\mbox{Si\,{\sc iv}}}

\newcommand{\FeII}{\mbox{Fe\,{\sc ii}}}

\newcommand{\HI}{\mbox{H\,{\sc i}}}
\newcommand{\HII}{\mbox{H\,{\sc ii}}}

\title[The XQR-30 quasar sample]{XQR-30: the ultimate XSHOOTER quasar sample at the reionization epoch}

\author[V. D'Odorico et al.]{
Valentina D'Odorico$^{1,2,3}$\thanks{E-mail: valentina.dodorico@inaf.it},
E. Ba\~{n}ados$^4$, G. D. Becker$^5$, M. Bischetti$^{1,6}$, S. E. I. Bosman$^{4}$, G. Cupani$^{1,3}$,
\and 
 R. Davies$^{7,8}$, E. P. Farina$^9$, A. Ferrara$^2$, C. Feruglio$^{1,3}$, C. Mazzucchelli$^{10}$, E. Ryan-Weber$^{7,8}$,
\and
J.-T. Schindler$^{4,11}$, A. Sodini$^1$, B. P. Venemans$^{11}$, F. Walter$^4$, H. Chen$^{12}$, S. Lai$^{13,14}$, Y. Zhu$^5$, 
\and
F. Bian$^{14}$, S. Campo$^1$, S. Carniani$^2$, S. Cristiani$^{1,3}$, F. Davies$^4$, R. Decarli$^{15}$, A. Drake$^{4}$, A.-C. Eilers$^{16}$\thanks{Pappalardo fellow}, 
\and
X. Fan$^{17}$, P. Gaikwad$^4$, S. Gallerani$^2$, B. Greig$^{18,8}$, M. G. Haehnelt$^{19}$, J. Hennawi$^{20,11}$, L. Keating$^{21}$, 
\and 
G. Kulkarni$^{22}$,   A. Mesinger$^2$, R. A. Meyer$^4$, M. Neeleman$^{23}$, M. Onoue$^{24,25,4}$, A. Pallottini$^2$, Y. Qin$^{18,8}$,  
\and
S. Rojas-Ruiz$^4$, S. Satyavolu$^{22}$, A. Sebastian$^{7,8}$, R. Tripodi$^{6,1,3}$, F. Wang$^{17}$, M. Wolfson$^{20}$, J. Yang$^{17}$, 
\and 
 M. V. Zanchettin$^{26,1}$ 
\\
$^{1}$INAF--Osservatorio Astronomico di Trieste, Via G.B. Tiepolo, 11, I-34143 Trieste, Italy \\
$^2$Scuola Normale Superiore, P.zza dei Cavalieri, I-56126 Pisa, Italy\\
$^3$IFPU--Institute for Fundamental Physics of the Universe, via Beirut 2, I-34151 Trieste, Italy \\
$^{4}$Max-Planck-Institut f\"ur Astronomie, K\"onigstuhl 17, D-69117 Heidelberg, Germany \\
$^5$Department of Physics \& Astronomy, University of California, Riverside, CA 92521, USA \\
$^6$Dipartimento di Fisica, Sezione di Astronomia, Universit\'a di Trieste, via Tiepolo 11, 34143 Trieste, Italy \\
$^7$Centre for Astrophysics and Supercomputing, Swinburne University of Technology, Hawthorn, Victoria 3122, Australia \\
$^8$ARC Centre of Excellence for All-Sky Astrophysics in 3 Dimensions (ASTRO 3D), Australia \\
$^9$Gemini Observatory, NSF’s NOIRLab, 670 N A’ohoku Place, Hilo, Hawai'i 96720, USA \\
$^{10}$Instituto de Estudios Astrof\'{\i}sicos, Facultad de Ingenier\'{\i}a y Ciencias, Universidad Diego Portales, Avenida Ejercito Libertador 441, Santiago, Chile\\
$^{11}$Leiden Observatory, Leiden University, P.O. Box 9513, 2300 RA Leiden, The Netherlands \\
$^{12}$Canadian Institute for Theoretical Astrophysics, University of Toronto, 60 St George St, Toronto, ON. M5S 3H8, Canada \\
$^{13}$Research School of Astronomy and Astrophysics, Australian National University, Canberra, ACT 2611, Australia \\
$^{14}$European Southern Observatory, Alonso de C\'ordova 3107, Casilla 19001, Vitacura, Santiago 19, Chile \\
$^{15}$INAF--Osservatorio di Astrofisica e Scienza dello Spazio di Bologna, via Gobetti 93/3, 40129, Bologna, Italy \\
$^{16}$MIT Kavli Institute for Astrophysics and Space Research, 77 Massachusetts Ave., Cambridge, MA 02139, USA \\
$^{17}$Steward Observatory, University of Arizona, 933 N Cherry Avenue, Tucson, AZ 85721, USA \\
$^{18}$School of Physics, University of Melbourne, Parkville, VIC 3010, Australia \\
$^{19}$Kavli Institute for Cosmology and Institute of Astronomy, Madingley Road, Cambridge, CB3 0HA, UK \\
$^{20}$Department of Physics, University of California, Santa Barbara, CA 93106, USA \\
$^{21}$Institute for Astronomy, University of Edinburgh, Blackford Hill, Edinburgh, EH9 3HJ, UK \\
$^{22}$Tata Institute of Fundamental Research, Homi Bhabha Road, Mumbai 400005, India \\
$^{23}$National Radio Astronomy Observatory, Charlottesville, USA \\
$^{24}$Kavli Institute for Astronomy and Astrophysics, Peking University, Beijing 100871, China \\
$^{25}$Kavli Institute for the Physics and Mathematics of the Universe (Kavli IPMU, WPI), The University of Tokyo, Chiba 277-8583, Japan \\ 
$^{26}$SISSA, Via Bonomea 265, I-34136 Trieste, Italy
}

\date{Accepted XXX. Received YYY; in original form ZZZ}

\pubyear{2015}

\begin{document}
\label{firstpage}
\pagerange{\pageref{firstpage}--\pageref{lastpage}}
\maketitle

\begin{abstract}
  The final phase of the reionization process can be probed by rest--frame UV absorption spectra of quasars at $z\gsim6$, shedding light on the properties of the diffuse intergalactic medium within the first Gyr of the Universe. The ESO Large Programme “XQR-30: the ultimate XSHOOTER legacy survey of quasars at $z \simeq 5.8-6.6$” dedicated $\sim250$ hours of observations at the VLT to create a homogeneous and high-quality sample of spectra of 30 luminous quasars at $z\sim6$, covering the rest wavelength range from the Lyman limit to beyond the \MgII\ emission. Twelve quasar spectra of similar quality from the XSHOOTER archive were added to form the enlarged XQR-30 sample, corresponding to a total of $\sim350$ hours of on-source exposure time. The median effective resolving power of the  42 spectra is $R \simeq 11400$ and $9800$ in the VIS and NIR arm, respectively. The signal-to-noise ratio per 10 \kms\ pixel ranges from $\sim11$ to 114 at $\lambda\simeq 1285$ \AA\ rest frame, with a median value of $\sim29$. We describe the observations, data reduction and analysis of the spectra, together with some first results based on the E-XQR-30 sample. New photometry in the $H$ and $K$ bands are provided for the XQR-30 quasars, together with composite spectra whose characteristics reflect the large absolute magnitudes of the sample.  The composite and the reduced spectra are released to the community through a public repository, and will enable a range of studies addressing outstanding questions regarding the first Gyr of the Universe.
\end{abstract}

\begin{keywords}
intergalactic medium - galaxies: high-redshift - quasars: absorption lines - quasars: emission lines - dark ages, reionization, first stars
\end{keywords}



\section{Introduction}
The first billion years of the Universe define the current frontier of modern observational cosmology. 
During this time the first stars and galaxies assembled from the primordial gas, and the atomic hydrogen permeating the early Universe became ionized. 
The epoch of reionization (EoR) represents a major phase transition in cosmic history, which impacted almost every baryon in the Universe. 
Over the last decade, a coherent picture of the Universe during the EoR has started to emerge. 
Numerical models that best match a range of independent observations, including many of those described below, now generally feature a mid-point of reionization near $z\sim7-8$ and an end well below $z\sim6$ \citep[e.g.][]{kulkarni2019,keating2020,nasir2020,choudhury2021,qin2021,cain2021,garaldi2022}. This suggests that much of the reionization process may lie within a redshift range that is highly accessible observationally. An integrated constrain on the EoR comes from the measurement of the optical depth, $\tau_{\rm es}$, of the Thomson scattering of cosmic microwave background photons on the free electrons. The most recent measurements \citep{Planck2018}, report a value $\tau_{\rm es} = 0.0561 \pm 0.0071$, which can be converted into a mid-reionization redshift $z_{\rm re} = 7.82\pm0.71$, in agreement with the previous picture.  

Quasars at $z \sim6$, due to their high intrinsic luminosities and prominent Lyman-$\alpha$ (\Lya) emission lines, have played a relevant role in the early characterization of the reionization process \citep[see][for a recent review]{fan2022}. The forest of \HI\ \Lya\ absorptions extending blueward of $\lambda \simeq 1216$ \AA\ in the quasar rest frame traces the diffuse intergalactic gas whose ionization is sensitive to the metagalactic ultraviolet background (UVB). 
As the \Lya\ opacity in the forest increases with redshift, eventually complete absorption is reached once the intergalactic medium (IGM) reaches average hydrogen neutral fractions of $\langle x_{\rm HI}\rangle \gsim 10^{-4}$ \citep{GP1965}. 

The rapid increase in the mean optical depth ($\tau_{\rm GP}$), together with the appearance of large regions of total absorption 
gave the first indication that reionization finished by redshift $z \sim 5-6$ \citep{fan2006,becker2015a}.
Other approaches have also been used to constrain the volume averaged \HI\ fraction with quasar spectra, most
notably: the IGM \HI\ damping wing in quasars at $z>7$ (e.g., \citealt{greig2017,greig2019,greig2022,davies2018,wang2020,yang2020a}), 
and the evolution of \HII\ proximity zones around larger collections of quasars (\citealp{maselli2009}; \citealp{carilli2010}; \citealp{venemans2015a}, but see also \citealp{eilers2017,davies2020} showing how the size of the proximity zone is independent of the neutral hydrogen fraction). 

Another class of constraints, based on the detail of high-$z$ \Lya\ transmission, includes: the ‘dark pixel’ fraction \citep{mesinger2010,mcgreer2015}, which provides the most model-independent measurement of the neutral fraction; the distribution function of totally absorbed dark gap lengths \citep{gallerani2008,malloy2015}, and the morphological analysis of individual spikes of transmission \citep{barnett2017,chardin2018,kakiichi2018,meyer2020,gaikwad2020,yang2020b} which are sensitive to the strength of the local ionization field, the clustering of sources, and IGM temperature. 

Moreover, the scatter in opacity between high-redshift quasar lines of sight probing the same redshift interval gave
the first evidence of an inhomogeneous UVB persisting until $z \sim 5.3$ \citep{becker2015a,bosman2018,eilers2018,yang2020b}. This observation gave rise to multiple models invoking e.g. early heating of the IGM by the first galaxies \citep{daloisio2018,keating2018}, the action of rare powerful sources \citep{chardin2017} and variations in the mean free path of ionizing photons originating from early galaxies \citep{davies2016}.  However, the observed sightline-to-sightline scatter is best reproduced with simulations assuming a late reionization, which has a midpoint at $z\sim7$ and completes by $z \simeq 5.3$ \citep[e.g.][]{kulkarni2019}. 

\begin{figure}
    \includegraphics[width=\columnwidth]{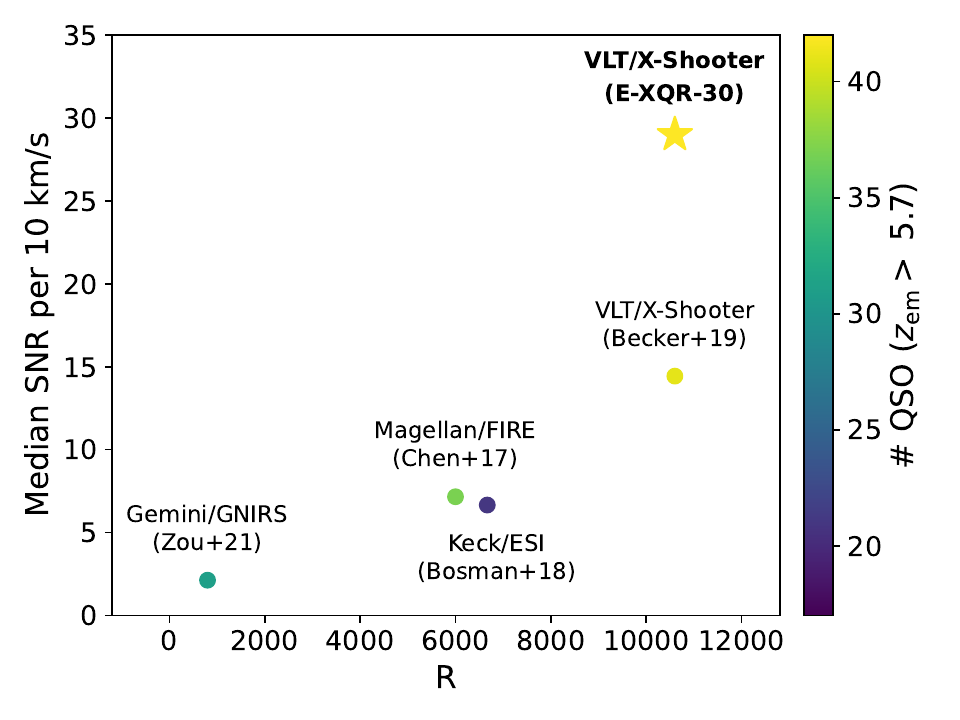}
    \caption{Comparison of the properties of E-XQR-30 (star) with other spectroscopic surveys of quasars at $z\sim6$ in terms of spectral resolution, median SNR and number of targets (color scale in the sidebar). See the text for more details. }
    \label{fig:survey_comp}
\end{figure}

As pointed out by some of these studies, these techniques are severely limited by the scarcity of spectroscopic data with a high signal-to-noise ratio (SNR).
Although the number of known quasars at $z\sim6$ has increased by more than a factor of 10 in recent years 
\citep[e.g.][]{banados2016,reed2019,wang2019,banados2022,matsuoka2022,yang2023}, robust conclusions on these important cosmological issues have to rely on the small available sample of $\sim10$ quasar spectra with high enough spectral resolution ($R \gsim 8000$) and SNR ($\gsim 10-20$ per 10 \kms\ pixel).
With the aim of making a decisive step forward in this research field, we conceived the {\it XQR-30} Large Programme (id. 1103.A-0817, P.I. V. D'Odorico) which provides a public sample of 30 high-SNR spectra of quasars at $5.8 \le \zem \le 6.6$ obtained with the XSHOOTER spectrograph \citep{vernet2011} at the ESO Very Large Telescope (VLT). XSHOOTER is a high-sensitivity, Echelle spectrograph at intermediate resolution ($R \approx 6000-9000$), providing complete wavelength coverage from the atmospheric cutoff to the NIR ($\sim0.3$ to 2.5 $\mu$m) in one integration. 

The ``enlarged" XQR-30 sample (E-XQR-30) comprises the XQR-30 targets and the other 12 high-quality quasar spectra in the same redshift range available in the XSHOOTER archive. 
Figure~\ref{fig:survey_comp} shows the unique properties of E-XQR-30 in terms of resolving power, SNR and number of targets when compared with those of other spectroscopic surveys of quasars with $z_{\rm em} > 5.7$ used for absorption line studies \citep{chen2017,bosman2018,becker2019,zou2021}. The comparison samples were selected among those published to date being obtained with a given instrument and with available  median SNR values. In particular, the sample of XSHOOTER spectra in \citet{becker2019} is a picture of the situation before the XQR-30 survey. We also note that XSHOOTER is the instrument that offers the broadest, simultaneous wavelength coverage with respect to Magellan/FIRE ($0.82-2.51$ $\mu$m), Gemini/GNIRS ($0.85-2.5$ $\mu$m) and Keck/ESI ($0.39-1.1$ $\mu$m).

The characteristics of this sample makes it ideal to investigate several topics in addition to allowing the study of the \HI\ reionization process. Within the XQR-30 collaboration, we focus on: 
\begin{itemize}
\item[$\bullet$] {\it the early chemical enrichment and the mechanisms by which heavy elements are expelled from galaxies into the circum-galactic medium (CGM) and the IGM.} Measurements of the number density and cosmic mass density\footnote{The cosmic mass density, $\Omega$, is defined as the ratio between the comoving mass density of a given ion, determined from the observed column density, and the critical mass density.} of \CIV\ ($\lambda\,\lambda$ 1548, 1551 \AA) and \SiIV\  ($\lambda\,\lambda$ 1394, 1403 \AA) \citep{ryanweber2009,simcoe2011,dodorico2013,codoreanu2018,meyer2019,dodorico2022} show a rapid increase of these quantities with decreasing redshift in the range $5\lsim z \lsim 6.3$, followed by a plateau to $z\sim1.5$. On the other hand, the same statistical quantities determined for low ionization lines reveal an almost constant behaviour in the whole redshift range \citep[\MgII\ $\lambda\,\lambda$ 2796, 2803 \AA,][]{chen2017}  or even a decrease between the highest redshift bin at $z \ge 5.7$ and lower redshifts \citep[\OI\ $\lambda$ 1302 \AA,][]{becker2019}. These results have intriguing implications for the evolution of the ionization state of the CGM and they have been found difficult to reproduce by cosmological simulations \citep[e.g.][]{keating2014,rahmati2016,keating2016,finlator2020,doughty2021}.   
However, the statistical significance of the results at $z \gsim 5$  is limited by the fact that they are based  on only a few tens of detected lines, and on spectral samples with heterogeneous characteristics.  
 
\item[$\bullet$] {\it the chemical abundances in the ISM of high-z galaxies and the nature of the first stars.} 
Damped \Lya\ systems\footnote{DLAs are absorption systems characterized by a neutral hydrogen column density N(\HI)$\ge 10^{20.3}$ cm$^{-2}$.} (DLAs) trace the kinematics and chemical composition of dense gas in and close to galaxies at $z \lsim 5$ \citep[e.g.][]{wolfe2005}. The extremely high \HI\ column density in these systems implies self-shielding from the external radiation. Metals are often dominated by a single ionization state and chemical abundances are determined with very high precision. This precision is also linked to the accuracy of the column density determinations that depends critically on the spectral resolution and on the SNR of the used spectra. At $z \gsim 5$, the increasing opacity of the IGM makes the direct identification of DLA troughs more and more difficult, requiring the use of other tracers. Luckily, several low ionization metal lines are still detectable in the region of the spectrum redward of the \Lya\  emission of the quasar \citep[see e.g.][]{cooper2019}. In particular, it is reasonable to assume that \OI\ absorption systems trace predominantly neutral gas, as the \OI\ ionization potential is similar to that of \HI, and are therefore the analogues of lower-redshift DLAs and sub-DLAs ($10^{19} < $ N(\HI)/cm$^{-2}$ $< 10^{20.3}$).  
To date, the relative abundances of O, C, Si, and Fe have been measured in 11 \OI\ systems at $z\sim 5 - 6$ \citep{becker2012,poudel2018,poudel2020}. In few cases, characterized by the proximity of the DLA to the quasar systemic redshift, it has been possible to measure the column density of \HI\ and derive the metallicity and absolute abundances of the systems \citep{dodorico2018,banados2019,andika2022}. 
Measured abundances at $z\gsim 5$ are generally consistent with those obtained in DLAs and sub-DLAs in the redshift range $2 < z < 4$ \citep[e.g.][]{dessauges2003,peroux2007,cooke2011}. The lack of strong variations in the absorption-line ratios suggests that these systems are enriched by broadly similar stellar populations. There is no clear evidence of unusual abundance patterns that would indicate enrichment from exotic sources such as metal-free Population III (PopIII) stars.

\item[$\bullet$] {\it the properties of quasars and their environment in the early Universe.}
The mere presence of luminous quasars, powered by fast accretion ($>10$ M$_{\sun}$ yr$^{-1}$) onto massive ($> 10^8$ M$_{\sun}$) black holes (BHs) less than 1 Gyr after the Big Bang, represents a challenge for models of massive BH formation and early galaxy growth \citep[e.g.][]{lupi2021}. Several mechanisms have been invoked for the formation of the seeds of these BHs \citep[e.g.][]{regan2009,latif2016,volonteri2021}, from the remnants of extremely massive PopIII stars, growing through phases of super-Eddington accretion, to more exotic scenarios involving the direct collapse of a single massive cloud of pristine gas. All these models yield distinctive predictions for the time scales of the BH mass growth, on the typical Eddington rate, on the BH to host galaxy stellar mass ratio, on the metallicity of the BH close environment, and on the galactic environment of the first quasars. These models can thus be tested by accurate BH mass and accretion rate measurements achieved via sensitive NIR spectroscopy of the \MgII\ line \citep[e.g.][]{shen2019,yang2021,farina2022}. The metallicity of the broad line region can also be reconstructed via the \MgII/\FeII\ ratio, that is used as a proxy of the abundance of $\alpha$ elements \citep[e.g.][]{derosa2011,mazzucchelli2017,onoue2020,schindler2020,yang2021,wang2022}. The measurements of \HII\ proximity zone sizes can be used to estimate quasar episodic lifetimes \citep{eilers2017,eilers2020} as well as duty cycles \citep{davies2020,satyavolu2023}, which can lead to constraints on the BH obscuration, seed mass and formation redshifts.
\end{itemize}

In this paper, we present the XQR-30 sample and its extension to all the available XSHOOTER spectra of $z \gsim 5.8$ quasars with similar quality. The target selection is described in Section~\ref{sec:selection}, the details of observations are reported in Section~\ref{sec:obs}. In Section~\ref{sec:reduction} we detail the steps of the data reduction. The procedures adopted for the absolute flux calibration, the computation of the effective spectral resolution and the intrinsic continuum determination are explained in Section~\ref{sec:analysis}. The E-XQR-30 data products released to the community are described in Section~\ref{sec:release}. Finally, we give a short summary of the results obtained up to now in Section~\ref{sec:result} and we draw our conclusions in Section~\ref{sec:conclusion}.     
Unless otherwise stated all reported magnitudes are given in the AB system. 
We adopt a $\Lambda$CDM cosmology with parameters $\Omega_{\Lambda} = 0.7$, $\Omega_{\rm m} = 0.3$, and $h = 0.7$.

\begin{figure}
    \includegraphics[width=\columnwidth]{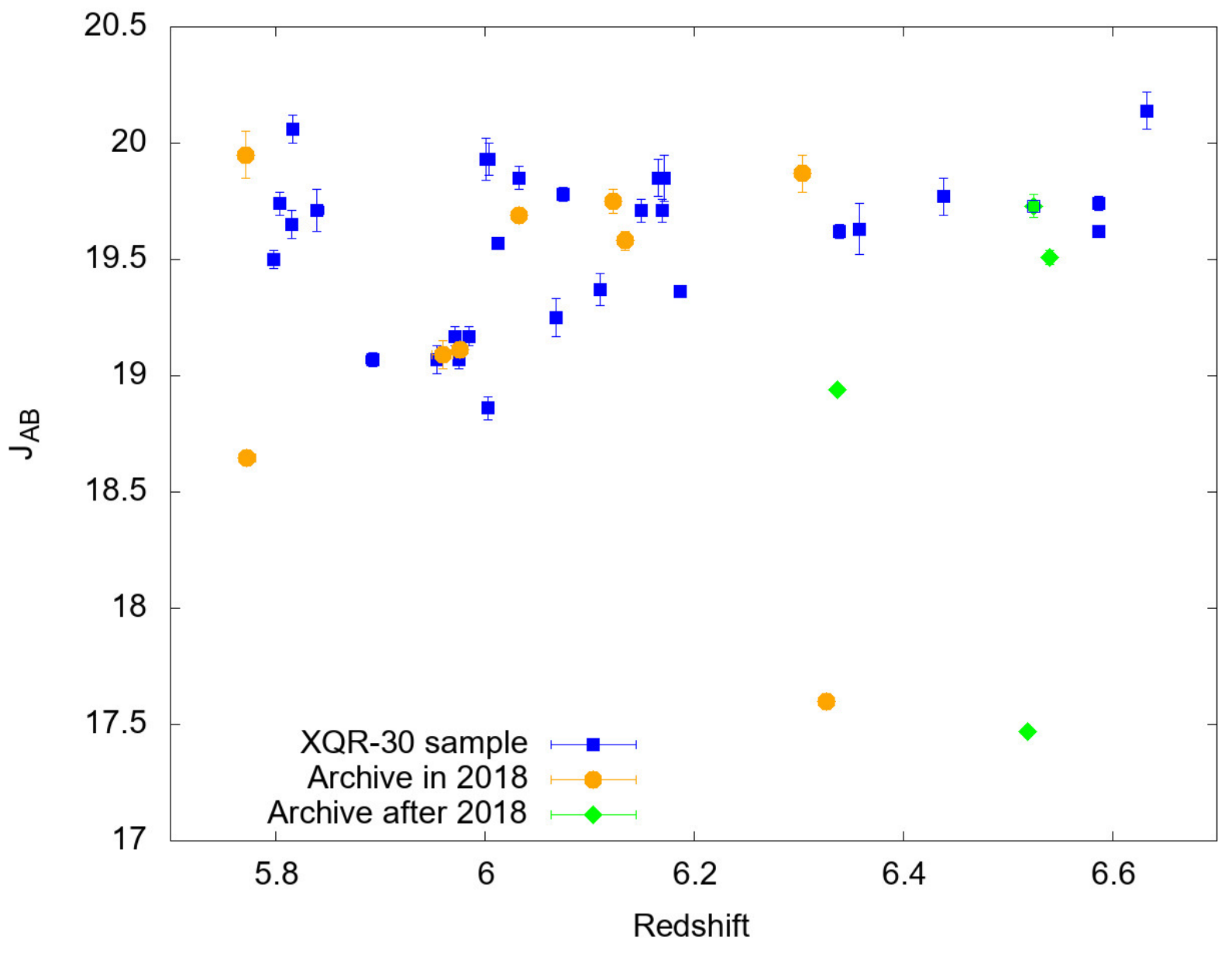}
     \caption{Distribution of the XQR-30 targets (blue squares) in the Redshift-$J_{\rm AB}$ magnitude plane. The orange dots represent the 9 quasars with SNR~$ \ge 25$ per pixel, already available from the XSHOOTER archive at the time of the proposal preparation. The green diamonds are the 4 quasars with similar properties whose spectra in the XSHOOTER archive became available after 2018. VDES J0224-4711 is represented as a green diamond with a blue contour (see text).}
     \label{fig:redshift_range}
\end{figure}

\begin{landscape}
\begin{table}
	\caption{Properties of the quasars in the enlarged XQR-30 sample (XQR-30 + literature quasar spectra with similar quality). See the text for the detailed description of the table columns. }
	\label{tab:xqr30_sample}
	\scriptsize 
	\begin{tabular}{lcccccccccccccc} 
		\hline
		name & RA (J2000) & DEC (J2000) & J$_{\rm AB}$ & Ref & H$_{\rm AB}$ & Ref. & K$_{\rm AB}$ & Ref &  $m_{1450}$ &  $z_{\rm [CII]}$ & Ref. & Dis.\ Ref.\  & $z_{\rm MgII}$ & log M$_{\rm BH}$  \\
  (1) & (2) & (3) & (4) & (5) & (6) & (7) & (8) & (9) & (10) & (11) & (12) & (13) & (14) & (15)\\
		\hline
PSO J007+04  &  00:28:06.56 & +04:57:25.64 & 19.93 $\pm$ 0.09 & 1$^a$ & 20.05 $\pm$ 0.03 & H 2021-09-12 &  19.79 $\pm$ 0.05 & S 2019-12-13 & 20.35 & 6.0015 $\pm$ 0.0002 & 8 & 1,5 & 5.954 $\pm$ 0.008 & 9.89$^{0.09}_{0.11}$ \\
PSO J009-10&  00:38:56.52 & -10:25:53.90 & 19.93 $\pm$ 0.07 & 2 & 19.55 $\pm$ 0.05 &  H 2021-09-12 &  19.24 $\pm$ 0.07 & S 2019-12-13  & 20.50 & 6.0040 $\pm$ 0.0003 & 8 & 2 & 5.938 $\pm$ 0.008 & 9.90$^{0.09}_{0.11}$\\  
PSO J023-02  &  01:32:01.70 & -02:16:03.11 & 20.06 $\pm$ 0.06 & 2$^a$ & 20.08 $\pm$ 0.04 & H 2021-09-12 & 19.70 $\pm$ 0.17 & 4 & 20.61 & -- & -- &  2 & 5.817 $\pm$ 0.002 &  9.39$^{0.05}_{0.06}$ \\
PSO J025-11  &  01:40:57.03 & -11:40:59.48 & 19.65 $\pm$ 0.06 & 2 & 19.78 $\pm$ 0.07 & H 2021-09-12 & 19.60 $\pm$ 0.04 & S 2019-12-13 & 20.02 & -- & -- & 2 & 5.816 $\pm$ 0.002 &  9.33$^{0.06}_{0.07}$ \\
PSO J029-29  &  01:58:04.14 & -29:05:19.25 & 19.07 $\pm$ 0.04 & 2 & 19.14 $\pm$ 0.02 & H 2021-09-12 & 18.91 $\pm$ 0.03 & 4  & 19.44 & -- & -- & 2 & 5.976 $\pm$ 0.003 &   9.34$^{0.06}_{0.07}$ \\
VST-ATLAS J029-36 &  01:59:57.97 & -36:33:56.60 & 19.57 $\pm$ 0.01 & 3 & 19.62 $\pm$ 0.02 & H 2021-09-12 & 19.23 $\pm$ 0.06 & 4 & 19.85 &  --  & -- & 7 & 6.013 $\pm$ 0.006 & 9.20$^{0.06}_{0.07}$ \\
VDES J0224-4711 & 02:24:26.54 & -47:11:29.40 & 19.73 $\pm$ 0.05 & 4 & 19.45 $\pm$ 0.07  & 4  & 18.94 $\pm$ 0.05 & 4 &  20.13 & -- & -- & 9 & 6.525 $\pm$ 0.001  & 9.19$^{0.07}_{0.08}$ \\
PSO J060+24  &  04:02:12.69 & +24:51:24.42 & 19.71 $\pm$ 0.05 & 2 & 19.72 $\pm$ 0.08 & H 2021-09-13 & 19.53 $\pm$ 0.06 & 4 &  20.03 & -- & -- & 2 & 6.170 $\pm$ 0.003 & 9.18$^{0.07}_{0.09}$ \\
VDES J0408-5632 &  04:08:19.23 & -56:32:28.80 & 19.85 $\pm$ 0.05 & 4 & 19.68 $\pm$ 0.05 & H 2021-09-12 & 19.63 $\pm$ 0.05 & S 2019-12-14 & 20.23 & -- & -- & 9 & 6.033 $\pm$ 0.004 & 9.31$^{0.08}_{0.09}$\\  
PSO J065-26  &  04:21:38.05 & -26:57:15.60 & 19.36 $\pm$ 0.02 & 3 & 19.15 $\pm$ 0.04 & H 2021-09-12 & 19.22 $\pm $0.06 & 4 & 19.60 & 6.1871 $\pm$ 0.0003 & 8 &2 &  6.179 $\pm$ 0.006 & 9.56$^{0.13}_{0.18}$ \\  
PSO J065+01  &  04:23:50.15 & +01:43:24.79 & 19.74 $\pm$ 0.05  & 3 &  19.48 $\pm$ 0.03 &  H 2021-09-13 & 19.37 $\pm$ 0.05 & S 2019-12-13 & 20.55 & -- & -- & 13 & 5.804 $\pm$ 0.004 & 9.60$^{0.12}_{0.17}$ \\
PSO J089-15  &  05:59:45.47 & -15:35:00.20 & 19.17 $\pm$ 0.04 & 2 & 18.78 $\pm$ 0.03 & H 2021-09-13  & 18.43 $\pm$ 0.04 & S 2019-12-13 & 19.74 &  -- & -- & 2 & 5.972 $\pm$ 0.005 & 9.57$^{0.08}_{0.10}$ \\
PSO J108+08  &  07:13:46.31 & +08:55:32.65 & 19.07 $\pm$ 0.06 & 4 & 19.03 $\pm$ 0.07 & 2 & -- &  -- &  19.57 & -- &  & 2 & 5.955 $\pm$ 0.002 & 9.49$^{0.08}_{0.10}$ \\
SDSS J0842+1218 &  08:42:29.43 & +12:18:50.58 & 19.78 $\pm$ 0.03 & 5 & -- & -- & 19.22 $\pm$ 0.04 & S 2019-12-12 & 20.18 & 6.0754 $\pm$ 0.0005 & 8 & 5 & 6.067 $\pm$ 0.002 &   9.30$^{0.07}_{0.09}$\\ 
DELS J0923+0402  &  09:23:47.12 & +04:02:54.40 & 20.14 $\pm$ 0.08 & 4 &  19.90 $\pm$ 0.13 & 4& 19.44 $\pm$ 0.07 & 4 & 20.32 & 6.6330 $\pm$ 0.0003 & 14   & 15 & 6.626 $\pm$ 0.002 & 9.42$^{0.16}_{0.24}$ \\
PSO J158-14  &  10:34:46.50 & -14:25:15.58 & 19.25 $\pm$ 0.08 & 4  & 19.15 $\pm$ 0.10 & 4 &  18.68 $\pm$ 0.09 & 4 & 19.72 & 6.0685 $\pm$ 0.0001 & 10 & 13,16 & 6.065 $\pm$ 0.002 & 9.31$^{0.07}_{0.08}$ \\
PSO J183+05  &  12:12:26.98 & +05:05:33.49 & 19.77 $\pm$ 0.08 & 6 & 19.54 $\pm$ 0.11 & 4 & 19.54 $\pm$ 0.11 & 4 & 19.98 & 6.4386 $\pm$ 0.0002 &8 & 6 & 6.428 $\pm$ 0.005 & 9.41$^{0.21}_{0.41}$ \\
PSO J183-12  &  12:13:11.81 & -12:46:03.45 & 19.07 $\pm$ 0.03 & 4 & 19.10 $\pm$ 0.04 & 4 & 19.10 $\pm$ 0.08 & 4 &  19.47 & -- & -- & 1 & 5.893 $\pm$ 0.006 & 9.22$^{0.07}_{0.09}$ \\
PSO J217-16  &  14:28:21.39 & -16:02:43.30 & 19.71 $\pm$ 0.05 & 4 & 19.73 $\pm$ 0.11 & 4 & 19.37 $\pm$ 0.09 & 4 &  20.09 & 6.1498 $\pm$ 0.0011 & 11 & 2 &  6.135 $\pm$ 0.003 & 9.02$^{0.19}_{0.34}$ \\
PSO J217-07  &  14:31:40.45 & -07:24:43.30 & 19.85 $\pm$ 0.08 & 4 & 19.99 $\pm$ 0.04 & H 2021-09-12 & 19.82 $\pm$ 0.15 & 4 & 20.37 & -- & -- & 2 & 6.166 $\pm$ 0.004 & 8.90$^{0.16}_{0.27}$ \\
PSO J231-20  & 15:26:37.84 & -20:50:00.66 &  19.62 $\pm$ 0.02 & S 2021-07-27 & 19.55 $\pm$ 0.005 & S 2021-07-27 & 19.30 $\pm$ 0.03 & S 2021-07-27  & 19.85 & 6.5869 $\pm$ 0.0004 & 8  & 6 & 6.564 $\pm$ 0.003 & 9.52$^{0.04}_{0.04}$\\ 
DELS J1535+1943  &  15:35:32.87 & +19:43:20.10  & 19.63 $\pm$ 0.11 & 4 & 19.01 $\pm$ 0.05 & H 2021-09-12 & -- & -- &  20.23 & --  & --  & 17 & 6.358 $\pm$ 0.002 & 9.80$^{0.06}_{0.07}$ \\
PSO J239-07  &  15:58:50.99 & -07:24:09.59 & 19.37 $\pm$ 0.07  & 4 & 19.27 $\pm$ 0.01 & H 2021-09-13  & 18.98 $\pm$ 0.09 & 4 & 19.74 & 6.1102 $\pm$ 0.0002 & 10 & 2 & 6.114 $\pm$ 0.001 & 9.42$^{0.05}_{0.06}$\\ 
PSO J242-12  &  16:09:45.53 & -12:58:54.11 & 19.71 $\pm$ 0.09 & 4 & 19.66 $\pm$ 0.02 & H 2021-09-14 & 19.46 $\pm$ 0.26 & 4 &  20.29 & -- & -- & 2 & 5.840 $\pm$ 0.006  & 9.51$^{0.10}_{0.13}$\\  
PSO J308-27 &   20:33:55.91 & -27:38:54.60 & 19.50 $\pm$ 0.04 & 2$^a$ & 19.62 $\pm$ 0.05 & H 2021-09-07  & 19.63 $\pm$ 0.15 & 4 &  19.95 & -- & -- & 2 & 5.799 $\pm$ 0.002  & 9.09$^{0.05}_{0.05}$ \\  
PSO J323+12  & 21:32:33.19 & +12:17:55.26 &  19.74 $\pm$ 0.03 & 6 & 19.65 $\pm$ 0.06 & H 2021-09-07 & 19.21 $\pm$ 0.02 & F 2017-09-24 & 20.06 & 6.5872 $\pm$ 0.0004 & 8 &6 & 6.585 $\pm$ 0.002 &  8.92$^{0.09}_{0.13}$ \\
VDES J2211-3206  &  22:11:12.17 & -32:06:12.94 & 19.62 $\pm$ 0.03 & 13 & 19.43 $\pm$ 0.04 & H 2021-09-12 & 19.00 $\pm$ 0.03 & 13 & 19.95 & 6.3394 $\pm$ 0.0010 & 11 & 13,16 & 6.330 $\pm$ 0.003 & 9.30$^{0.15}_{0.24}$\\
VDES J2250-5015  &  22:50:02.01 & -50:15:42.20 & 19.17 $\pm$ 0.04 & 4 & 19.12 $\pm$ 0.05 &  H 2021-09-12 & 18.94 $\pm$ 0.05 & 4 &  19.80 & -- & -- & 9 & 5.985 $\pm$ 0.003 & 9.66$^{0.41}_{0.41}$ \\
SDSS J2310+1855 &  23:10:38.89 & +18:55:19.70 &  18.86 $\pm$ 0.05 & 4 & 18.94 $\pm$ 0.04 & N 2020-07-12 & 18.96 $\pm$ 0.05 & N 2020-07-12 & 19.26 & 6.0031 $\pm$ 0.0002 & 12 & 18 & 5.992 $\pm$ 0.003 & 9.67$^{0.06}_{0.08}$ \\
PSO J359-06  &  23:56:32.45 & -06:22:59.26 & 19.85 $\pm$ 0.10 & 2 & 19.70 $\pm$ 0.02 & H 2021-09-12 & 19.32 $\pm$ 0.02 & F 2017-09-25 & 20.2 & 6.1722 $\pm$ 0.0001 & 10 & 2 & 6.169 $\pm$ 0.002 & 9.00$^{0.09}_{0.12}$ \\ 
\hline
SDSS J0100+2802 & 01:00:13.02 & +28:02:25.80 & 17.60 $\pm$ 0.02 & 4 & 17.48 $\pm$ 0.02 & 4 & 17.04 $\pm$ 0.16 & 21 & 17.70 & 6.3269 $\pm$ 0.0002 & 8 &  21 & 6.316 $\pm$ 0.003 & 10.09$^{0.21}_{0.39}$ \\
VST-ATLAS J025-33   & 01:42:43.72 & -33:27:45.61 & 18.94 $\pm$ 0.01 & F 2017-09-02 & 18.95 $\pm$ 0.02 & 4 & 18.76 $\pm$ 0.03 &1& 19.07 & 6.3373 $\pm$ 0.0002 & 8 & 7 & 6.330 $\pm$ 0.003 & 9.37$^{0.17}_{0.27}$ \\ 
ULAS J0148+0600 & 01:48:37.64 & +06:00:20.06 & 19.11 $\pm$ 0.02 & 4 & 19.12 $\pm$ 0.07 & 4  & 19.02 $\pm$ 0.08 & 4 & 19.49 & -- & --  & 7 & 5.977 $\pm$ 0.002 & 9.58$^{0.08}_{0.10}$ \\
PSO J036+03     & 02:26:01.87 & +03:02:59.24 & 19.51 $\pm$ 0.03 & 19 & 19.30 $\pm$ 0.09 & 4 & 19.30 $\pm$ 0.09 &4& 19.65 & 6.5405 $\pm$ 0.0001 &8 &  19 & 6.527 $\pm$ 0.004 & 9.43$^{0.08}_{0.10}$ \\
WISEA J0439+1634     & 04:39:47.08 & +16:34:16.01  & 17.47 $\pm$ 0.02 & 4 & 17.33 $\pm$ 0.17 & 20 & 16.85 $\pm$ 0.01 & 4& 17.74 & 6.5188 $\pm$ 0.0004 & 14 &  20 & 6.520 $\pm$ 0.002 & 9.72$^{0.07}_{0.09}$ \\
SDSS J0818+1722 & 08:18:27.40  & +17:22:52.01 &  19.09 $\pm$ 0.06 & 4 & -- & -- & -- & -- & 19.56 & -- & -- &  23 &  5.960 $\pm$ 0.010 &  9.76$^{0.06}_{0.08}$\\
SDSS J0836+0054 & 08:36:43.86 & +00:54:53.26  & 18.647 $\pm$ 0.003 & 4 & 18.878 $\pm$ 0.004 & 4 & 18.355 $\pm$ 0.009 & 4 & 19.01 &  -- & --  & 24 & 5.773 $\pm$ 0.008 & 9.59$^{0.13}_{0.20}$ \\
SDSS J0927+2001 & 09:27:21.82 &+20:01:23.64  & 19.95 $\pm$ 0.10 & 20 & -- & -- & --  &-- & 20.26 & 5.7722 $\pm$ 0.0006 & 22$^b$  &  23  & 5.760 $\pm$ 0.005 & 9.11$^{0.10}_{0.13}$ \\
SDSS J1030+0524 & 10:30:27.11 & +05:24:55.06  & 19.87 $\pm$ 0.08 & 4 & 19.79 $\pm$ 0.11 & 4 & 19.49 $\pm$ 0.07 &4 & 20.20 &  -- & --   &24 & 6.304 $\pm$ 0.002 &  9.27$^{0.09}_{0.12}$ \\
SDSS J1306+0356 & 13:06:08.26 & +03:56:26.19 & 19.69 $\pm$ 0.03 & 4 & 19.65 $\pm$ 0.05 & 4 & 19.34 $\pm$ 0.04 & 4 & 19.98 & 6.0330 $\pm$ 0.0002 & 8 &24 & 6.020 $\pm$ 0.002 & 9.29$^{0.09}_{0.12}$ \\ 
ULAS J1319+0950 & 13:19:11.29 & +09:50:51.49 & 19.58 $\pm$ 0.04 & 4& 19.71 $\pm$ 0.13 & 4 & 19.44 $\pm$ 0.13 & 4 & 19.84 & 6.1347 $\pm$ 0.0005 & 8 & 25 & 6.117 $\pm$ 0.002 & 9.53$^{0.05}_{0.05}$  \\
CFHQS J1509-1749 & 15:09:41.78 & -17:49:26.80 & 19.75 $\pm$ 0.05 & 4 & --  & -- & 19.49 $\pm$ 0.09 & 4 & 20.15 & 6.1225 $\pm$ 0.0007 & 11 & 26  & 6.119 $\pm$ 0.003 & 9.30$^{0.15}_{0.22}$ \\
		\hline
	\end{tabular}
	\\
	References: 1 -  \citet{banados2014}; 2 - \citet{banados2016}; 3 - \citet{bischetti2022}; 4 - \citet{ross2020}; 5 - \citet{jiang2015}; 6 - \citet{mazzucchelli2017}; 7 - \citet{carnall2015}; 8 -  \citet{venemans2020}; 9 - \citet{reed2017}; 10 - \citet{eilers2021}; 11 - \citet{decarli2018}; 12 - \citet{wang2013}; 13 - \citet{banados2022}; 14 - \citet{yang2021}; 15 - \citet{matsuoka2018b}; 16 - \citet{chehade2018}; 17 - \citet{wang2019b}; 18 - \citet{jiang2016}; 19 - \citet{venemans2015a}; 20 - \citet{fan2019}; 21 - \citet{wu2015}; 22 - \citet{carilli2007}; 23 -  \citet{fan2006a}; 24 - \citet{fan2001}; 25 - \citet{mortlock2009}; 26 -  \citet{willott2007} \\
$^a$ Photometry has been re-analysed and updated in this paper. $^b$ Redshift from a CO transition.
\end{table}
\end{landscape}

\section{Sample selection}
\label{sec:selection}
We have selected the best objects available  to pursue our goals. 
In particular, our targets were the quasars with the brightest $J$ magnitude and redshift $z \ge 5.8$ known in 2018 and were chosen to properly sample the interesting redshift range (see Fig.~\ref{fig:redshift_range}). Our sample was selected from published quasars \citep[e.g.][]{banados2014,jiang2015,carnall2015,banados2016,reed2017} as well as newly discovered quasars, unpublished at the time of the proposal \citep{banados2022}. 
The targets were selected to have the following characteristics:

\begin{itemize}
\item{} observability from Paranal: $\delta < +27$ deg;
\item{} emission redshift range $5.8 \le \zem \le 6.6$ (corresponding to a cosmic time interval of $\sim150$ Myr). The lower $z$ limit is set by \MgII\ emission entering the K-band, the higher redshift by the requirement of a homogeneous redshift coverage due to the very limited number of quasars known above $z\sim6.5$ at the time;
\item{} magnitude cut J$_{\rm AB} \le 19.8$ for $\zem < 6.0$ and J$_{\rm AB} \le 20.0$ for $6.0 \le \zem \le 6.6$, based on the photometry known at the time of the proposal\footnote{Note that with the updated photometry reported in Table~\ref{tab:xqr30_sample} two objects no longer satisfy these limits: PSO J023-02 and J0923+0402. They have been kept in the sample anyway.};
\item{} no existing XSHOOTER data with SNR~$ \gsim 25$ per 10 \kms\ pixel.
\end{itemize}

Three objects, compliant with our target requirements, were not included in the XQR-30 Large Programme (LP) because they were already part of an accepted proposal by the same P.I. (id. 0102.A-0154, green diamonds with $J_{\rm AB} > 18.5$ in Fig.~\ref{fig:redshift_range}). However, one of these quasars, VDES J0224-4711, was eventually added to the XQR-30 sample and its observations were completed in the context of the LP (green diamond with blue contour in Fig.~\ref{fig:redshift_range}). The other two objects were included in the literature sample described in Section~\ref{sec:literat_sample}.

All the targets in the XQR-30 sample are listed in Table~\ref{tab:xqr30_sample}.  The columns of the table report: the quasar id. (1) and its spatial coordinates (2,3); the updated photometry (4-9) in $J$, $H$ and $K$ bands with the corresponding references or instrument used (S=SOFI, H=HAWK-I, F=Fourstar, N=NOTCam); the AB magnitude at 1450 \AA\ rest-frame measured from the XSHOOTER spectra (10); the emission redshift determined from the [\CII] emission line and the reference papers (11,12); the reference to the quasar discovery papers (13). The last two columns (14, 15) present the redshift based on \MgII\ \citep[][and this work]{bischetti2022} and the BH mass in M$_{\sun}$ measured from the \MgII\ FWHM (Mazzucchelli et al. 2023, A\&A submitted).
When available, we adopt the redshift determined from the [CII] 158 $\mu$m line in the rest frame Far-Infrared (FIR) of the observed quasars, which is a reliable tracer of the host galaxy systemic redshift \citep[e.g.][]{venemans2016,decarli2018}. The redshift determined from the \MgII\ emission is systematically lower than the one derived from [\CII] (with one exception) with velocity shifts ranging between $\sim50$ and 3000 \kms. At the redshift of our quasars, the [\CII] line is observable in the sub-mm wavelength regime; a companion ALMA programme (P.I. B. Venemans and S. Bosman) will obtain [\CII] and dust continuum observations for all the XQR-30 objects which were lacking this information. The complete list of [\CII] based systemic redshifts and the comparison with other emission lines will be published in a further paper (Bosman et al. in prep.).  

\section{Observations}
\label{sec:obs}

\subsection{XQR-30 objects}

The observations for the XQR-30 LP were carried out in “service mode” between April 7, 2019, and March 8, 2022. During this time XSHOOTER was mounted on the unit telescope (UT) 2 of the VLT until the end of February 2020, then it was moved to UT 3 and recommissioned in October 2020. Due to the COVID pandemic, there had been an interruption of operations at the ESO observatories from March 23 to October 21 2020, which delayed the execution of all observing programmes.  

Service mode allows the user to define the Observing Blocks, which contain the instrument setup and are carried out by the observatory under the requested weather conditions. 
The constraint set adopted for XQR-30 observations required a maximum airmass of 1.5, a fraction of lunar illumination $< 0.4$  and minimum moon distance of 60 degrees. 
The seeing constraint was set to 1.0'' (note that seeing is measured at 5000 \AA). ESO Large Programmes are granted high-priority status, which means that observations out of specifications are repeated and eventually carried over to the following semester (which is not always the case for Normal Programmes) until the constraints are met (to within $\approx10$\,\%). 

\begin{table}
	\centering
	\caption{Indicative exposure times adopted for the different magnitudes to reach the desired signal-to-noise ratio. }
	\label{tab:integr_time}
	\begin{tabular}{lc} 
		\hline
		J$_{\rm AB}$ & T$_{\rm exp} (h)$ \\
		\hline
		${\rm J} < 19.3$  & 4.0 \\
		$19.3 \le {\rm J} < 19.6$ & 6.0 \\
		$19.6 \le {\rm J} <19.8$ & 8.0 \\
		$19.8 \le {\rm J} < 19.9$ & 9.3 \\
		$19.9 \le {\rm J} < 20$ & 10.7 \\
		\hline
	\end{tabular}
\end{table}

XSHOOTER has three spectroscopic arms, UVB, VIS and NIR, covering the wavelength ranges $0.3 - 0.56$ $\mu$m, $0.55 - 1.05$ $\mu$m and $0.98 - 2.48$ $\mu$m, respectively. Each arm has its own set of shutter, slit mask, cross-dispersive element, and detector. The three arms are observed simultaneously. We aimed at obtaining an average signal-to-noise ratio  SNR~$\sim 25$ per 10 \kms\ pixel  as homogeneously as possible across the optical and near IR wavelength range and across the sample. Furthermore, we had as a goal to reach  a minimum SNR~$\sim8-10$  also in the crossing region between the VIS and NIR arm, where the efficiency decreases due to the dichroic.  This region, corresponding to the wavelength range $\sim 0.95-1.05$ $\mu$m, is critical for our goals since it covers the redshift ranges  $z \simeq 5.1-5.8$ for \CIV\ and $z \simeq 5.8-6.5$ for \SiIV. 

To reach the desired SNR, we split the sample into 5 magnitude bins for which we foresaw different total exposure times obtained summing individual integrations of 1200s (see Table~\ref{tab:integr_time}). 
Exposures were acquired nodding along the slit by $\pm2.5$'' from the slit center. This operation is carried out in order to improve the sky subtraction, in particular in the NIR wavelength range, which is performed by subtracting one from the other the pairs of nodded frames. However, as it is explained in Section~\ref{sec:reduction}, thanks to the very good performances of our custom pipeline, we were able to subtract the sky from every frame individually, improving the SNR.

The adopted slit widths were 0.9'' in the VIS and 0.6'' in the NIR arm, to account for the seeing wavelength dependence. These slit widths provide a nominal resolving power of 8900 and 8100, respectively. 

The UVB arm is not relevant for these high-redshift targets since the flux is completely absorbed below the Lyman limit (corresponding to $\lambda = 6200$ and 6930 \AA\ for $z=5.8$ and 6.6, respectively). The slit position was always set along the parallactic angle. 

Target acquisition was done in the SDSS $z'$ filter. However, since our targets are generally faint, the blind-offset technique was adopted, which points the telescope to a bright star close to the target and then move the telescope "blindly" to the position of the object to be observed. The VIS CCD was binned by a factor of 2 in the dispersion and spatial direction.
For each exposure, the standard calibration plan of the observatory was adopted. 

Some of the objects in our sample already had XSHOOTER observations in the ESO archive, which were included to generate our final spectra. The details of observations of each target are reported in Table~\ref{tab:obs_details}.

\begin{table*}
\footnotesize
	\centering
	\caption{Details of observations for the XQR-30 and literature samples. Exposure times (columns 2 and 3) are in hours and slit widths (columns 4 and 5) are in arcsec. Column 6 displays the ESO programme IDs of the reduced frames, while in column 7 and 8 the effective resolving power (see Sec.~\ref{sec:resol}) is reported; the SNR per 10 \kms\ bin measured in each final extracted spectrum at $\lambda\simeq 1285$ \AA\ rest frame is in column 9. Column 10 refers to the first publication of the considered XSHOOTER spectrum.    }
	\label{tab:obs_details}
	\begin{minipage}{17.5cm}
	\begin{tabular}{lcclllcccc} 
		\hline
Target     &    T$_{\rm VIS}$ & T$_{\rm NIR}$ & slit VIS & slit NIR & Programme ID & $R_{\rm VIS}$ & $R_{\rm NIR}$ & SNR & XS Ref. \\
\hline
PSO J007+04       &       9.7  &  9.0 &  0.9 & 0.6 & 098.B-0537, 1103.A-0817 & 11500 & 9700 & 28.5 & This work\\
PSO J009-10       &      10.3 &  10.7 &  0.9 & 0.6 & 097.B-1070, 1103.A-0817 & 11500 & 10500 & 23.9 & This work\\
PSO J023-02       &       9.3  &  9.3 &  0.9 & 0.6 & 1103.A-0817 & 10900 & 9700 & 18.8  & This work\\
PSO J025-11       &       9.3 &   9.3 &  0.9 & 0.6 & 1103.A-0817 & 10500 & 9700 & 27.7 & This work \\
PSO J029-29       &       4.0 &   4.0 &  0.9 & 0.6 & 0101.B-0272, 1103.A-0817 & 10800 & 9900 & 26.6 & This work \\
VST-ATLAS J029-36 &   7.4 &   7.5 &  0.9 & 0.9JH, 0.6 & 294.A-5031, 1103.A-0817 & 10100  & 9200 & 25.5& This work \\
VDES J0224-4711   &       8.6  &  8.3 &  0.9 & 0.9, 0.6 & 0100.A-0625, 0102.A-0154, 1103.A-0817 & 11200 & 9400 & 15.1 & This work\\
PSO J060+24       &       8.0 &  10.3 &  0.9 & 0.6 & 1103.A-0817 & 11500 & 10300 & 25.3 & This work \\
VDES J0408-5632       &      10.7  & 10.7 &  0.9 & 0.6 & 1103.A-0817 & 11300 & 9700 & 37.9 & This work \\
PSO J065-26       &       6.0  &  6.0 &  0.9 & 0.6 & 098.B-0537, 1103.A-0817 & 11700 & 10500 & 41.4 & This work \\
PSO J065+01       &      10.7 &  10.7 &  0.9 & 0.6 & 1103.A-0817  & 10700 & 9700 & 27.5 & This work \\
PSO J089-15       &       4.0  &  4.0 &  0.9 & 0.6 & 1103.A-0817  & 12000 & 9700 & 28.9 & This work\\
PSO J108+08       &       4.0  &  4.0 &  0.9 & 0.6 & 1103.A-0817  & 12200 & 9800 & 36.2 & This work \\
SDSS J0842+1218   &       8.7  &  8.7 &  0.9 & 0.6 & 097.B-1070, 1103.A-0817 & 11400 & 10000 & 37.5 & This work \\
DELS J0923+0402       &      12.0  & 12.0 &  0.9 & 0.6 & 1103.A-0817 & 11600 & 10200 & 11.0  & This work\\ 
PSO J158-14       &       5.8  &  5.9 &  0.9 & 0.9JH, 0.6 & 096.A-0418, 1103.A-0817 & 10800 & 9100 & 29.7 & This work\\
PSO J183+05       &       9.3  &  9.3 &  0.9 & 0.6 & 098.B-0537, 1103.A-0817 & 11900 & 10200 & 21.9  & This work\\
PSO J183-12       &       4.0  &  4.0 &  0.9 & 0.6 & 1103.A-0817 & 11900 & 10000 & 33.9 & This work\\
PSO J217-07       &      11.3 &  11.3 &  0.9 & 0.6 & 1103.A-0817 & 11200 & 9900 & 24.2  & This work\\
PSO J217-16       &      10.3 &  10.7 &  0.9 & 0.6 & 1103.A-0817 & 11500 & 10200 & 35.5  & This work\\
PSO J231-20       &      12.2 &  12.7 &  0.9 & 0.6 & 097.B-1070, 1103.A-0817 & 10800 & 9800 &17.5  & This work\\
DELS J1535+1943       &       8.7  &  9.3 &  0.9 & 0.6 & 1103.A-0817  & 11600 & 9700 & 15.2  & This work\\
PSO J239-07       &       6.0 &   6.0 &  0.9 & 0.6 & 0101.B-0272, 1103.A-0817 & 11300 & 11000 & 34.5 & This work \\
PSO J242-12       &       8.8 &   8.7 &  0.9 & 0.6 & 1103.A-0817 & 10700 & 9700 & 25.4  & This work\\
PSO J308-27       &       7.0 &   7.0 &  0.9 & 0.6 & 0101.B-0272, 1103.A-0817 & 11800 & 10600 & 31.2 & This work\\
PSO J323+12       &       8.9 &   9.7 &  0.9 & 0.6 & 098.B-0537, 1103.A-0817 & 10900 & 9800 & 15.7 & This work \\
VDES J2211-3206       &       4.8  &  4.8 &  0.9 & 0.9, 0.6 & 096.A-0418, 098.B-0537, 1103.A-0817 & 10600 & 9100 & 12.7 & This work \\
VDES J2250-5015       &       3.3  &  4.0 &  0.9 & 0.6 & 1103.A-0817 & 10800 & 9600 & 17.9 & This work\\
SDSS J2310+1855   &       4.0  &  4.0 &  0.9 & 0.6 & 098.B-0537, 1103.A-0817 & 12700 & 9800 & 40.4  & This work\\
PSO J359-06       &      12.0 &  12.0 &  0.9 & 0.6 & 098.B-0537, 1103.A-0817 & 11300 & 10000 & 35.7 & This work \\
\hline
SDSS J0100+2802    &     11.0  & 11.0 &  0.9 & 0.6JH, 0.6 & 096.A-0095 & 11400 & 10300 & 102.1 & 1\\
VST-ATLAS J025-33 &   4.7  &  4.8 &  0.9 & 0.9JH, 0.6 & 096.A-0418, 0102.A-0154 & 11200 & 9300 & 29.5 & This work\\
ULAS J0148+0600    &     11.0  & 11.0 &  0.7 & 0.6 & 084.A-0390 & 13300 & 10000 & 59.9 & 2 \\
PSO J036+03       &       6.5  &  6.7 &  0.9 & 0.9, 0.6 & 0100.A-0625, 0102.A-0154 & 10700 & 9200 & 18.5 & 3\\
WISEA J0439+1634       &      14.9  & 15.2 &  1.2 & 0.9 & 0102.A-0478  & 9500 & 8200 & 113.9 & This work\\
SDSS J0818+1722    &      9.2  &  9.4 &  0.9, 1.5 & 0.9, 1.2 & 084.A-0550, 086.A-0574, 088.A-0897 & 11000 & 7600 & 52.3 & 1\\
SDSS J0836+0054    &      2.3  &  2.3 &  0.7 & 0.6 & 086.A-0162 & 13100 & 10200 & 37.2 & 1\\
SDSS J0927+2001    &     12.0  & 10.8 &  0.7, 1.5 & 0.6, 1.2 & 0.84.A-390, 088.A-0897 & 12900 & 9400 & 36.5 & 1\\
SDSS J1030+0524    &      7.6  &  7.1 &  0.9, 1.5 & 0.9 & 084.A-0360, 086.A-0162, 086.A-0574,  & 12300 & 8400 & 16.8 & 1\\
 & & & & & 087.A-0607 & & & & \\
SDSS J1306+0356    &     13.4  & 11.0 &  0.9, 0.7 & 0.9, 0.6 & 60.A-9024, 084.A-0390 & 12000 & 9600 & 33.4 & 1\\
ULAS J1319+0950    &     10.5  & 10.0 &  0.7 & 0.6 & 084.A-0390 & 13700 & 9800 & 40.9 & 2\\
CFHQS J1509-1749   &      8.3  &  6.9 &  0.9 & 0.9, 0.9JH & 085.A-0299, 091.C-0934 & 11800 & 8000 & 27.3 & 1\\
\hline 
	\end{tabular}
	\\
	References: 1 - \citet{bosman2018};  2 - \citet{becker2015a}; 3 - \citet{schindler2020} .
	\end{minipage}
\end{table*}

\subsection{Additional objects}
\label{sec:literat_sample}

The XQR-30 sample was complemented with all the other quasars fitting the same brightness and redshift selection criteria  and having an XSHOOTER spectrum  with a SNR comparable with that of our sample. Currently, there are 12 quasars with these characteristics in the ESO archive which are also reported in Table~\ref{tab:xqr30_sample} and in Fig.~\ref{fig:redshift_range} (orange dots and green diamonds). The details of their observations are listed in Table~\ref{tab:obs_details}. 
The literature sample was reduced with the same pipeline used for the main sample and described in Section~\ref{sec:reduction}.   

We refer to the  sample comprising the XQR-30 quasars plus the literature quasars as the {\it enlarged XQR-30 sample} (E-XQR-30).   The total integration time for the full dataset is of  $\sim350$ hours. 

\subsection{Photometry of XQR-30 quasars}
\label{sec:photometry} 
We compile NIR  $J$, $H$ and $K$ AB magnitudes of the E-XQR-30 objects in Table~\ref{tab:xqr30_sample}. We use the photometry to characterize the spectral energy distribution of the quasars and to calibrate their spectra (see Section~\ref{subsec:fluxcal}).  

For a significant fraction of the quasars, this NIR photometry already existed in the literature (references in Table~\ref{tab:xqr30_sample}). We have obtained follow-up photometry for objects missing this information or in cases where the published photometry had large uncertainties.  We obtained $JHK$ images with the SOFI Imaging Camera on the NTT telescope \citep{moorwood1998}, $JK$ images with Fourstar at Las Campanas Observatory \citep{persson2013}, $HK$ images with the NOTCam camera at the Nordic Optical Telescope\footnote{\url{http://www.not.iac.es/instruments/notcam/}}, and $H$-band observations with the HAWK-I imager at the VLT \citep{2004SPIE.5492.1763P,2008A&A...491..941K}.  The on-source exposure times for SOFI and Fourstar observations range between 5 to 10 min. The exposure time for $H$ and $K$ NOTCam observations were 18 and 27 min, respectively. The HAWK-I observations were part of a bad-weather filler program with exposure times of 12.5 min. 

The images were reduced with standard procedures: bias subtraction, flat fielding, sky subtraction, and stacking. The photometric zero points were calibrated against stellar sources in the UKIDSS \citep{lawrence2007}, VHS \citep{mcmahon2013}, or 2MASS \citep{2006AJ....131.1163S} NIR surveys. We report the follow-up photometry and their observation dates in Table~~\ref{tab:xqr30_sample}.




\begin{figure*}
    \centering
    \includegraphics[width=15.5cm]{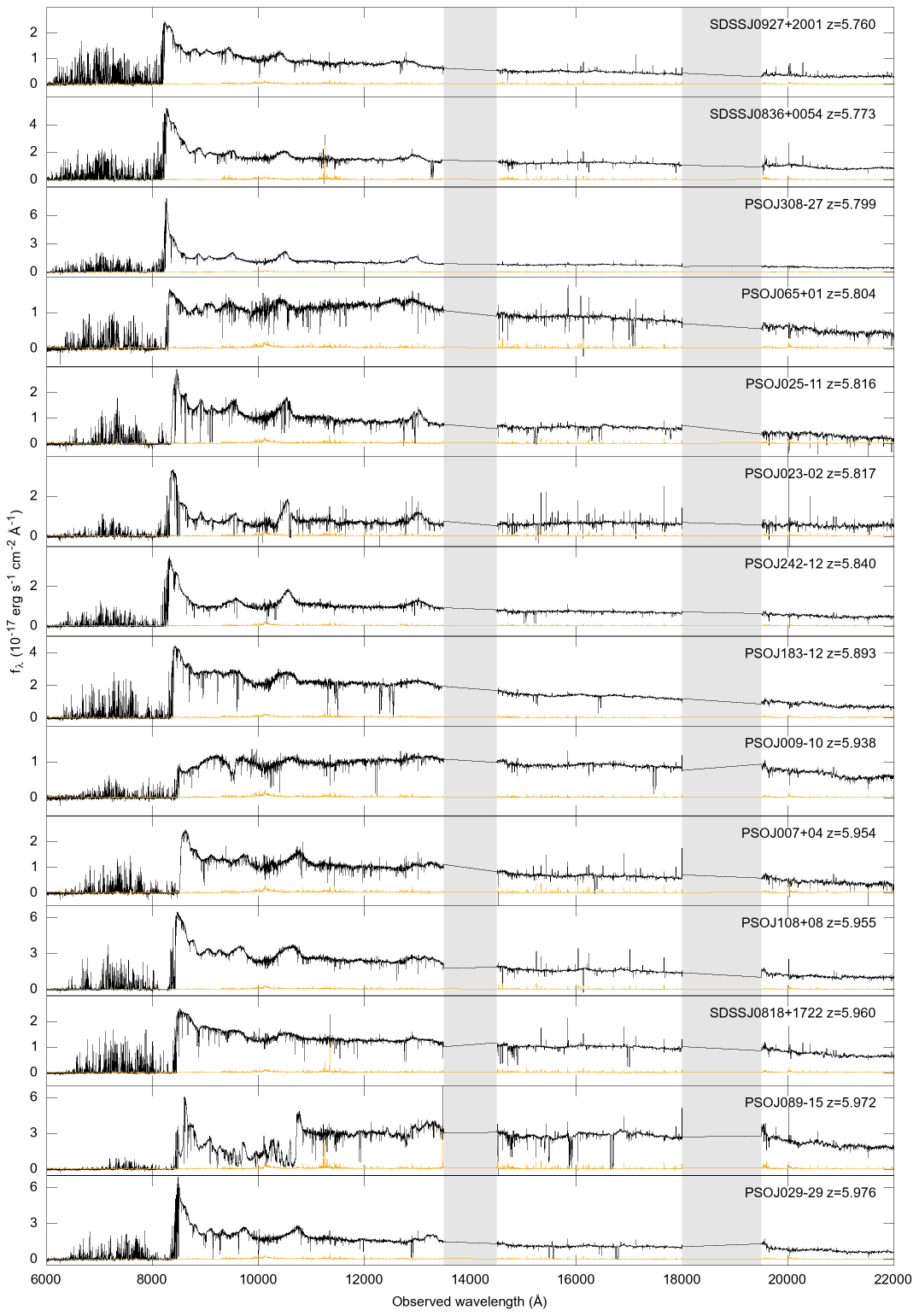}
     \caption{Spectra of the 42 E-XQR-30 quasars ordered by increasing \MgII\ redshift (as reported in Table~\ref{tab:xqr30_sample}). Each subplot reports the spectrum (black) and the error (yellow) arrays from 6000 to 22,000 \AA\ observed wavelengths. The grey bands correspond to the regions affected by strong telluric absorptions in the NIR, where the spectra have been cut. Spectra were rebinned to 50 \kms\ for display purposes.   }
     \label{fig:spectra_1}
\end{figure*}

\begin{figure*}
    \addtocounter{figure}{-1}
    \centering
    \includegraphics[width=16cm]{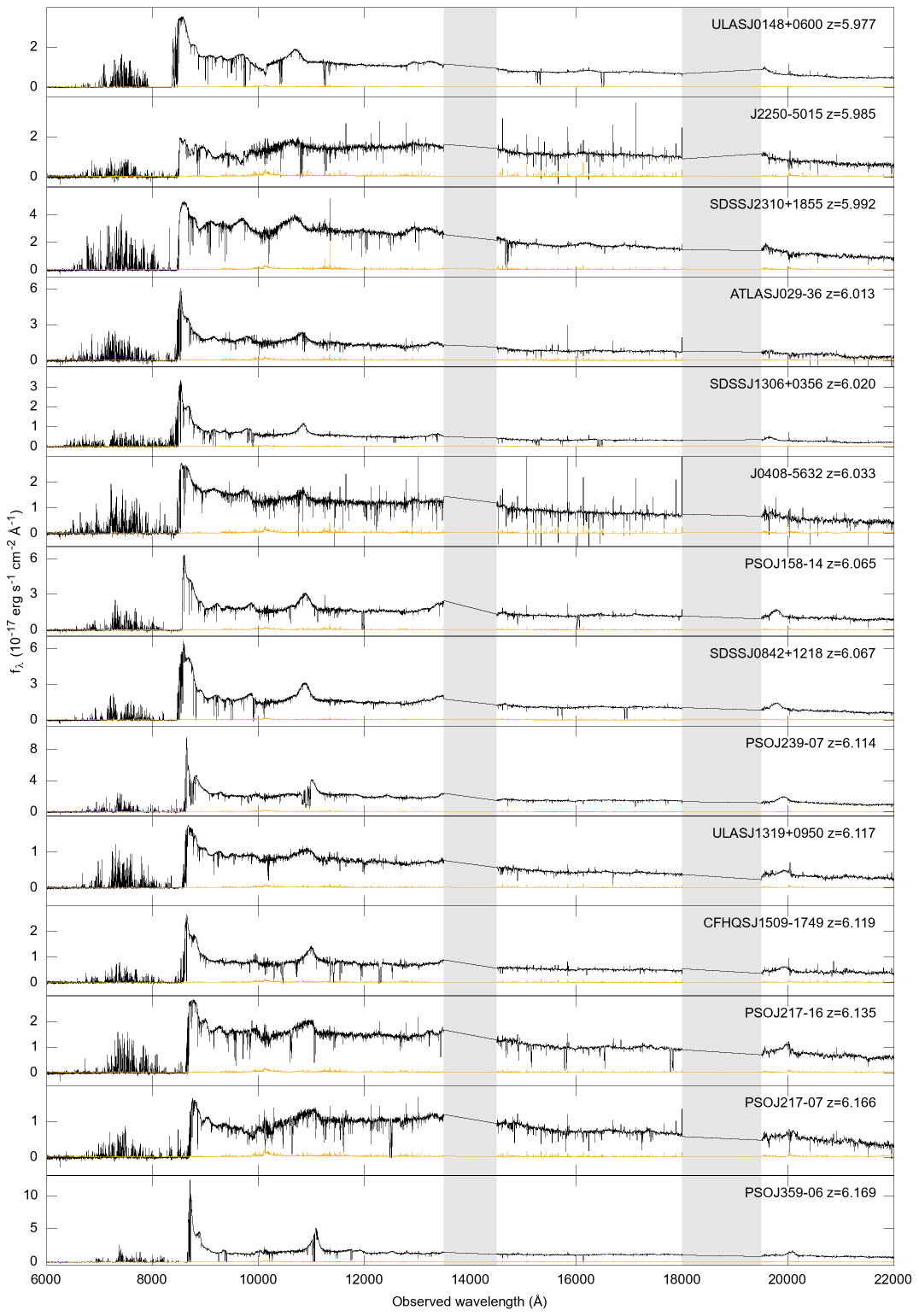}
     \caption{(Continued)}
     \label{fig:spectra_1}
\end{figure*}

\begin{figure*}
     \addtocounter{figure}{-1}
    \centering
     \includegraphics[width=16cm]{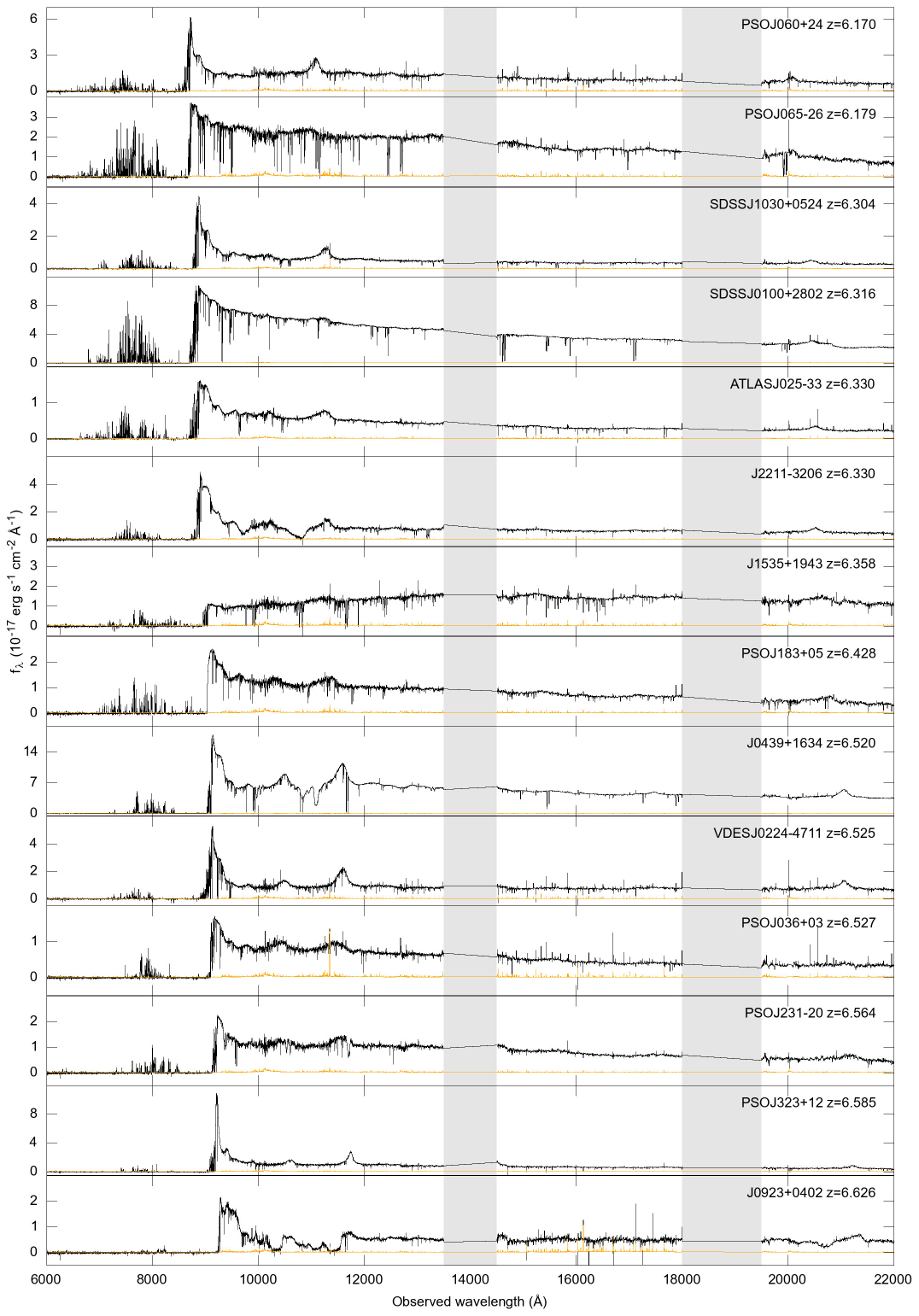}
     \caption{(Continued)}
     \label{fig:spectra_1}
\end{figure*}

\section{Data reduction}
\label{sec:reduction}

The XSHOOTER spectroscopic data were reduced using a modified version of the custom pipeline described in \citet{becker2012,becker2019} and \citet{lopez2016}.  The pipeline features highly optimized sky subtraction and spectral extraction, which perform better than 
the public XSHOOTER pipeline distributed by ESO, in particular for the sky subtraction in the NIR region \citep[for a comparison see][]{lopez2016}.  
The custom pipeline was applied to all new and archival frames included in the E-XQR-30 sample.  

Individual frames were first bias subtracted for the VIS arm and dark subtracted for the NIR.  A key feature of the pipeline is the use of high-SNR dark frames created by averaging many (generally 10--20) individual darks with exposure times matched to the science frames.  We typically used darks taken within a month of the science data for this purpose.  Subtracting these darks (along with flat-fielding) removes many of the detector features that often complicate sky subtraction in the near-infrared, allowing us to model and subtract the sky from each NIR frame individually rather than using nod-subtraction.  This avoids the factor of $\sqrt{2}$ penalty from nod-subtraction in the sky noise.  Individual frames were then flat-fielded.  Two-dimensional fits to the sky in the un-rectified frames were then performed following \citet{kelson2003} using a b-spline in the dispersion direction modified by a low-order polynomial along the slit.  A higher-order correction to the slit illumination function was also performed using the residuals from bright skylines.  

Spectral extraction and telluric correction were performed in multiple steps.  First, a preliminary one-dimensional spectrum was extracted from each frame.  Optimal extraction \citep{horne1986} was used adopting a Gaussian spatial profile, except for orders where the SNR was sufficient to fit a spline profile. Relative flux calibration was performed using a response function generated from a standard star.  Next, a fit to the telluric absorption in each frame was performed using models based on the Cerro Paranal Advanced Sky Model \citep{noll2012,jones2013}.  In these fits, the airmass was set to the value recorded in the header, while the precipitable water vapor, instrumental full width at half maximum (FWHM), and velocity offset between the model and spectrum were allowed to vary.  The best-fitting model was then propagated back to the two-dimensional sky-subtracted frame.  Finally, a single one-dimensional spectrum was obtained by applying the optimal extraction routine simultaneously to all exposures of a given object.  This last step allows for optimal rejection of cosmic rays and other artefacts.

All the reduced spectra are shown in Fig.~\ref{fig:spectra_1} and are publicly released (see Sec.~\ref{sec:release}). The SNR of the final extracted spectra measured at $\lambda \simeq 1285$ \AA\ rest frame is reported in Table~\ref{tab:obs_details}, the median value of the sample is SNR~$\sim 29$ per pixel of 10 \kms. 

\begin{figure}
     \includegraphics[width=\columnwidth]{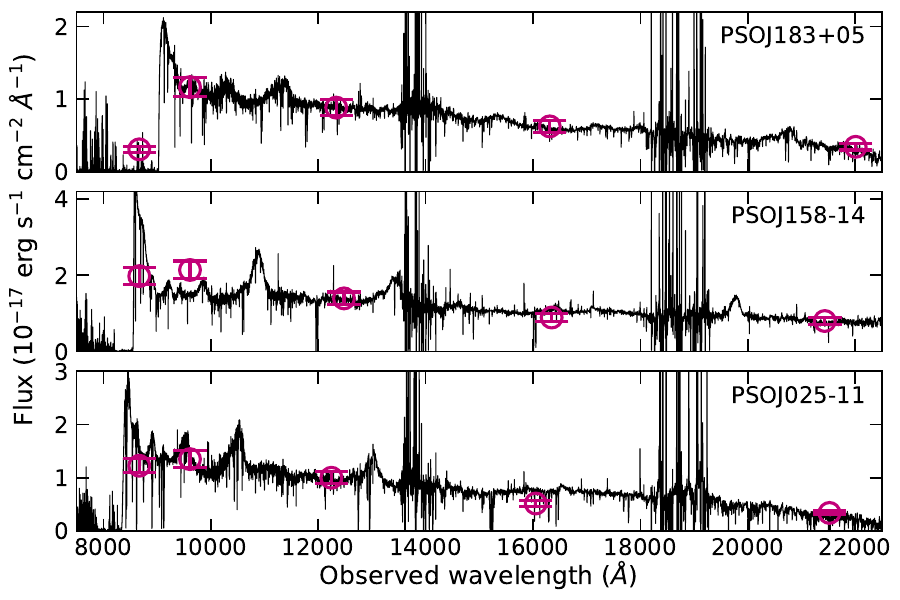}
     \caption{Three examples of XQR-30 quasar spectra, rebinned to 50 \kms\ and normalized to the J magnitude, compared with the other available photometric bands. In the upper panel the calibrated spectrum is in agreement with the photometry in the other bands; in the central panel, we show an object for which the $y$ band shows an excess, while the lower panel displays a spectrum with deficient H band photometry.}
     \label{fig:flux_norm}
\end{figure}

\section{Data Analysis}
\label{sec:analysis}

\subsection{Absolute flux calibration} 
\label{subsec:fluxcal}
The flux calibration of the spectra carried out by the data reduction pipeline is relative, in the sense that the shape of the spectrum is corrected for the instrument response but the flux values at the different wavelengths may differ from the true flux. This could be due to slit losses and non-photometric sky conditions at the time of observations. 

However, some of the science cases based on the E-XQR-30 sample do require an absolute flux calibration. To this aim, we created combined optical $+$ NIR spectra with the \textsc{Astrocook} python software \citep{cupani2022} through the following steps: 
\begin{itemize}
\item the NIR arm spectrum was scaled to the VIS one using the median values computed in the overlapping spectral region $1.00-1.02$ $\mu$m; 
\item the red end of the VIS arm spectrum and the blue end of the NIR arm were trimmed at $\lambda = 1.01$ $\mu$m to eliminate the noisiest regions;
\item the two spectra were stitched together and rebinned to 50 \kms. 
\end{itemize}

The combined spectrum was then normalized to the $J_{\rm AB}$ band photometry, assuming that the shape of the spectrum is correct \citep[see][]{bischetti2022}. We tested this assumption by verifying that the spectral shape does not significantly change when using the custom pipeline described above and the standard ESO XSHOOTER pipeline \citep{freudling2013}. To this aim, we considered the ESO-released reduced spectra which are calibrated with standard stars acquired on the same night of the observations. 

We note that the $J$-band photometry is not taken simultaneously with the spectra, which could introduce uncertainty in the absolute flux calibration, if some of these quasars have varied between the photometric and spectroscopic observations. 
However, there is an anti-correlation between quasar variability and luminosity or accretion rate (e.g. \citealp{sanchez2018}, but see also \citealp{kozlowski2019}). Given that the E-XQR-30 quasars are among the most luminous quasars at any redshift and accreting close to the Eddington limit, we do not expect that quasar variability would have a strong role in our results. 

The median absolute flux correction is $20\%$. As a further check, we also compared the $z$, $y$\footnote{$z$ and $y$ magnitudes for the E-XQR-30 sample are taken mainly from \citet{banados2016,banados2022} and are not reported in this paper. }, $H$ and $K$ photometry with the fluxes extracted from the normalized spectrum in the corresponding bands. In general, we find good agreement, with differences smaller than the $2\sigma$ photometric uncertainties. For two quasars, namely PSO J239-07, and PSO J158-14, the $y$ band photometry shows a flux excess by about 40\% with respect to the spectrum-based value. Finding differences between the $y$-band fluxes from Pan-STARRS1 (PS1) and the flux-calibrated spectra is not surprising, since the red side of the bandpass of the PS1 $y$-band is mainly determined by the CCD quantum efficiency (QE) and the PS1 camera has 60 CCDs (or more accurately, Orthogonal Transfer Arrays devices) all with different QEs \citep{tonry2012}. In three quasars (PSO J308-27, PSO J025-11 and PSO J023-02), we  observe a deficiency in the $H$ band by about 30\% ( Fig.~\ref{fig:flux_norm}). The largest $H$-band discrepancy of about 50\% is observed for SDSS J0836+0054. 

We computed the Galactic extinction for our quasars based on the SFD extinction map \citep{schlegel1998} recalibrated with the 14\% factor by \citet{schlafly2011}. The average (median) value for our sample is generally low, $E(B-V) \simeq 0.07$ (0.04). However, there are three quasars for which $E(B-V) \gsim 0.2$ (PSO J060+24, PSO J089-15 and PSO J242-12) and one (WISEA J0439+1634) with $E(B-V) \simeq 0.51$. 
Since the effect on the flux is typically negligible at the observed wavelengths, the released spectra are not corrected for Galactic extinction.

\subsection{Determination of effective spectral resolution}
\label{sec:resol}

The nominal resolving power of the VIS and NIR arms of XSHOOTER depend on the choice of the slit width. However, if the seeing at the time of the observation is smaller than the slit, the resolving power of the observed spectrum will be larger than the nominal one.  

An estimate of the resolving power as accurate as possible is critical for the process of Voigt profile fitting of the absorption lines. To this end we have investigated the observables linked to the spectral resolution in order to find an objective procedure to determine the effective resolving power of the spectra. 
These quantities are: 
\begin{itemize}
    \item the average FWHM seeing condition (in arcsec, at 5000 \AA) calculated over the exposure time as measured by the Differential Image Motion Monitor (DIMM) station at Paranal, which is reported in the ESO archive for each observed frame;
    \item the FWHM (in arcsec) of a Gaussian fit of the spectral order spatial profiles fitted in the 2D frames. This value depends on the position in the frame and, in the VIS arm, it can be determined only for the red orders since moving toward bluer wavelengths the flux is almost completely absorbed. We considered the FWHM at  $\sim9500$ \AA\ and 11,900 \AA\ for the VIS and the NIR arm, respectively;
    \item the FWHM (in \kms) of the Gaussian kernel by which the telluric model was smoothed in order to match the data. In the NIR spectrum, five FWHM are measured over the wavelength intervals: [10,900$-$12,850], [12,850$-$13,495], [14,830$-$16,240], [16,240$-$17,890] and [19,770$-$20,700] \AA\ in order to check whether the FWHM depends on wavelength. At least in the high SNR frames, no dependence has been detected to within $\sim 1$ \kms. Thus for the purpose of measuring the spectral resolving power, we take the mean value of these determinations. 
\end{itemize}

\begin{figure}
    \includegraphics[width=\columnwidth]{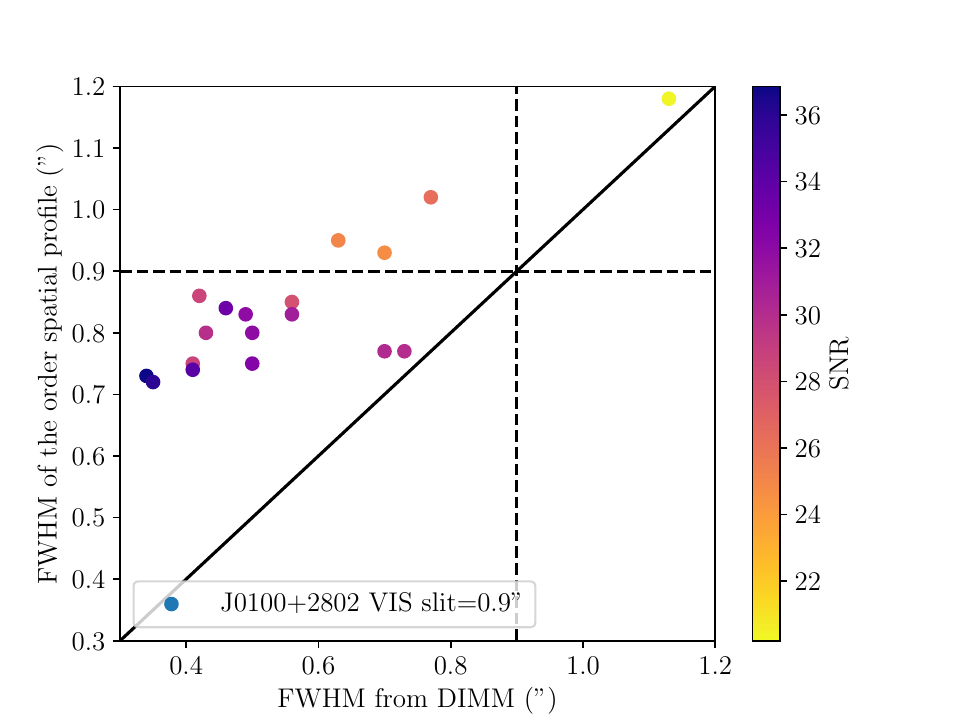}
	\includegraphics[width=\columnwidth]{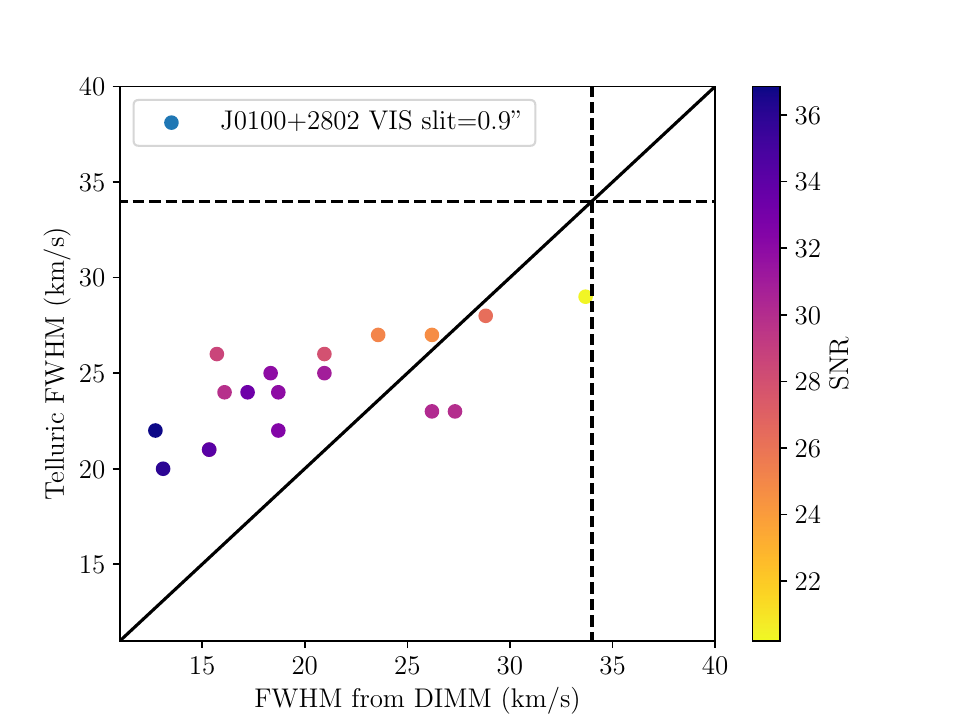}
    \caption{FWHM of the spectral order spatial profiles ({\it upper panel}) and average FWHM of the telluric model ({\it lower panel}) as a function of the FWHM at $\sim950$ nm of the seeing disk for the VIS frames of the literature quasar SDSSJ0100+2802,  which was observed with a slit=0.9". The dots are coloured according to the SNR of the corresponding frame (see the scale in the sidebar). The dashed lines indicate the slit value.  }
    \label{fig:resol_VIS_J0100}
\end{figure}

\begin{figure}
    \includegraphics[width=\columnwidth]{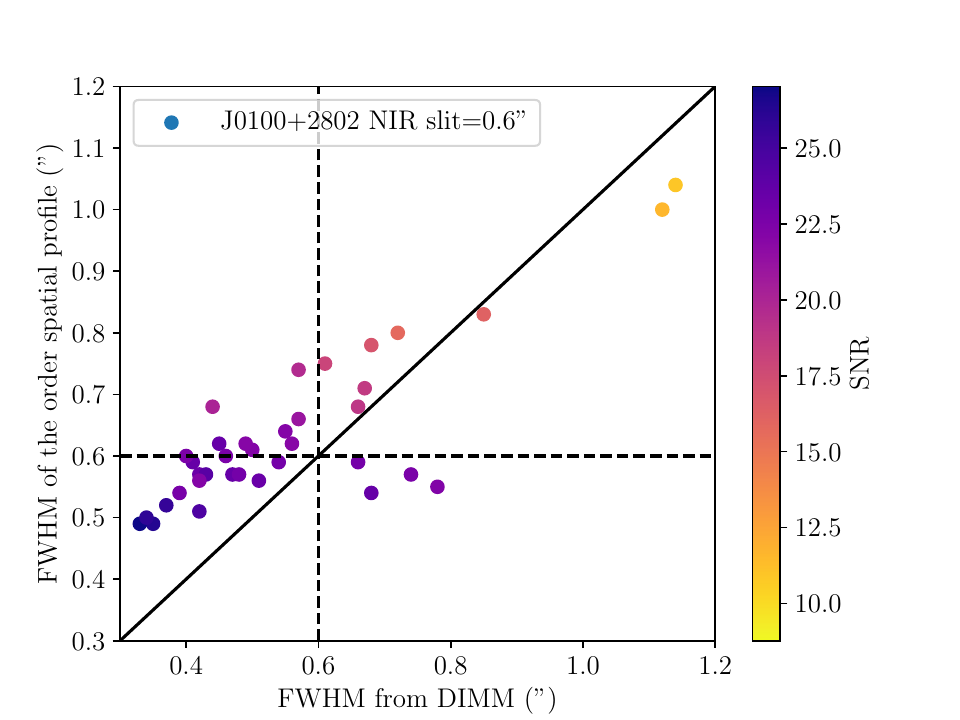}
	\includegraphics[width=\columnwidth]{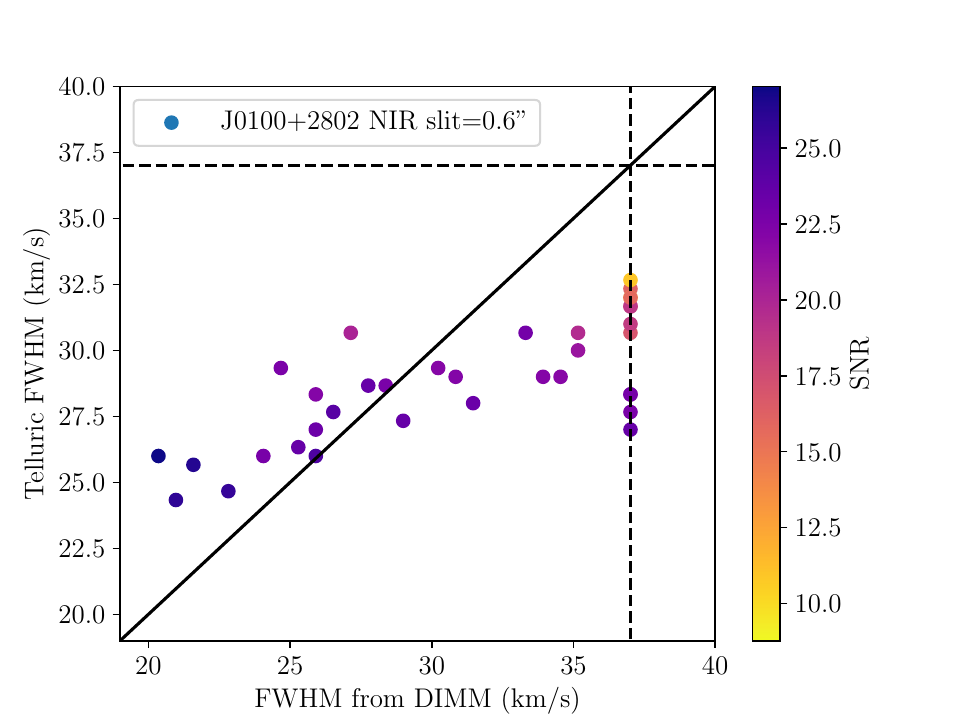}
    \caption{Same as Fig.~\ref{fig:resol_VIS_J0100} but for the  NIR frames of SDSSJ0100+2802, observed with a slit=0.6".  }
    \label{fig:resol_NIR_J0100}
\end{figure}

\begin{figure}
    \includegraphics[width=\columnwidth]{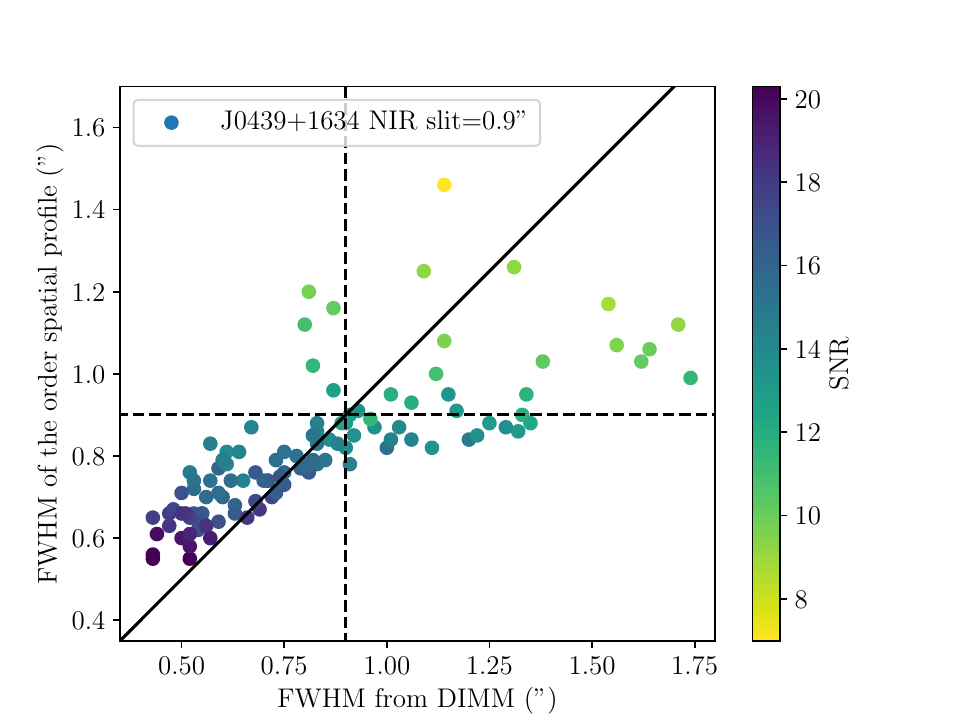}
	\includegraphics[width=\columnwidth]{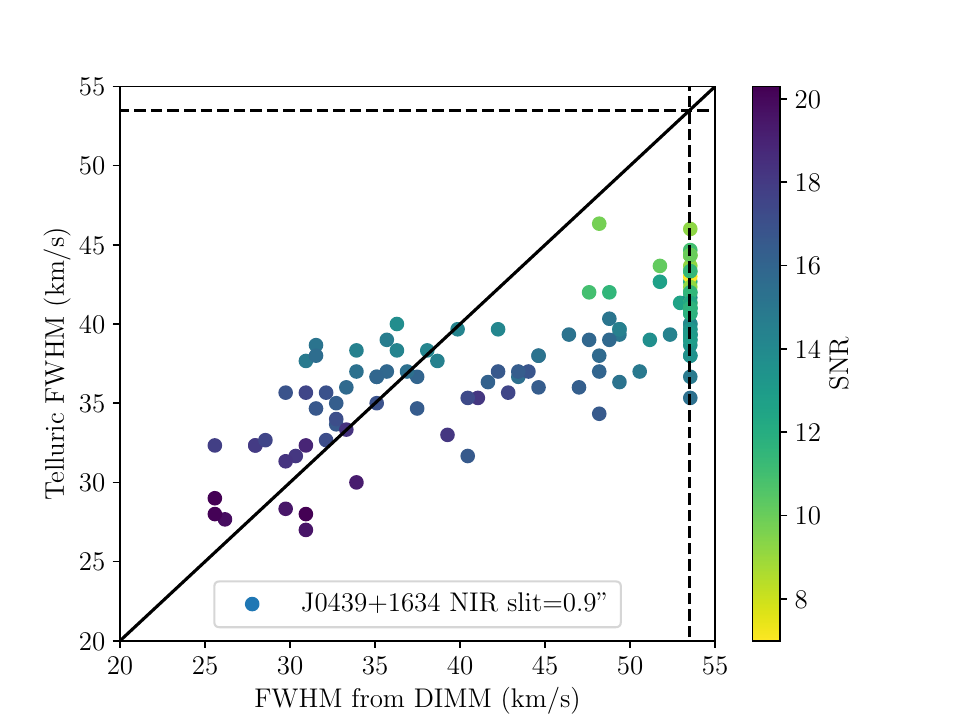}
    \caption{Same as Fig.~\ref{fig:resol_VIS_J0100} but for the  NIR frames of the literature quasar J0439+1634, observed with a slit=0.9". }
    \label{fig:resol_NIR_J0439}
\end{figure}

\begin{figure}
    \includegraphics[width=\columnwidth]{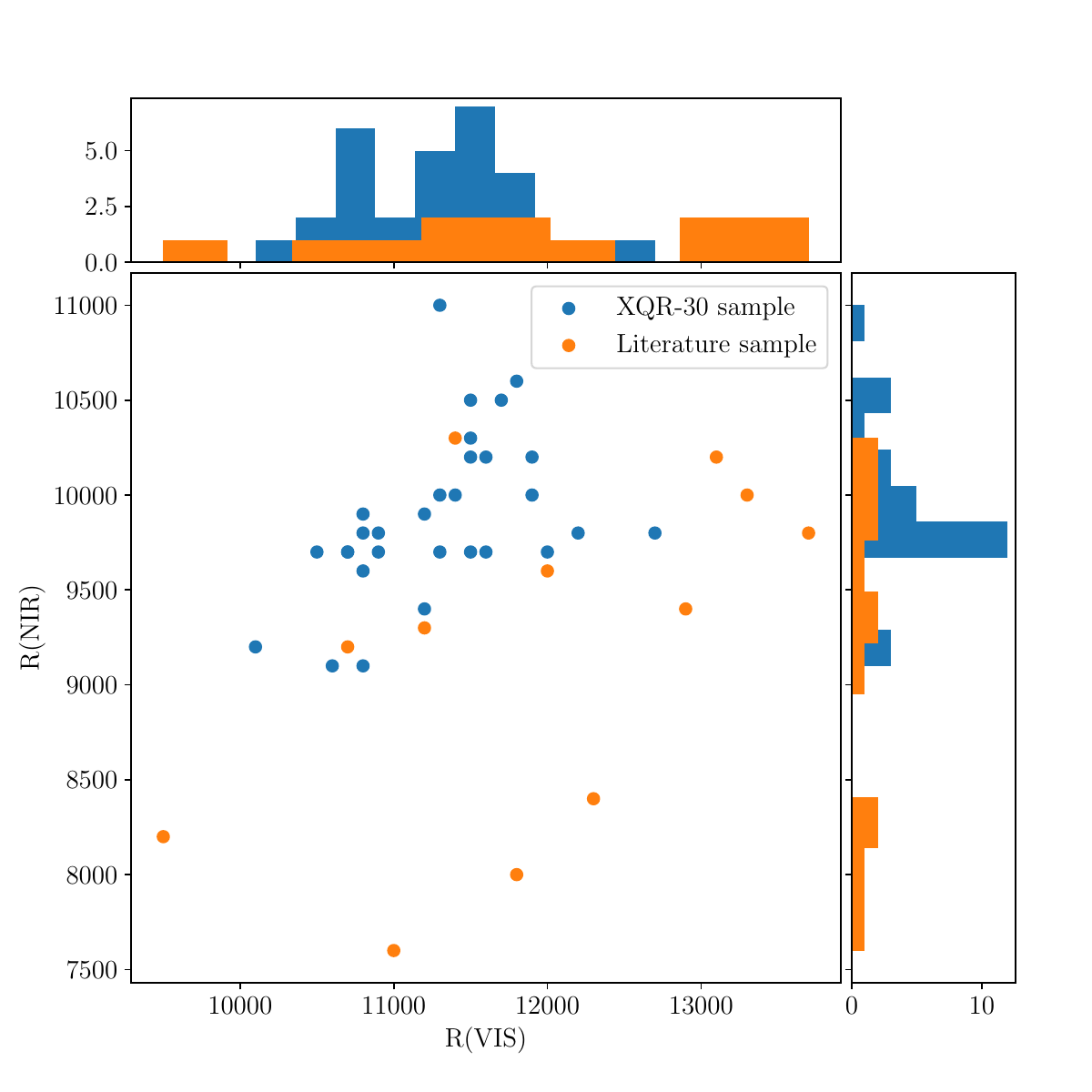}
    \caption{Resolving power in the NIR arm compared wit that in the VIS arm and relative distribution histograms for the XQR-30 (blue dots) and literature (orange dots) samples.   }
    \label{fig:resol_vs_resol}
\end{figure}

In Fig.~\ref{fig:resol_VIS_J0100}, we plot the relations between the above quantities for the VIS frames of the quasar SDSSJ0100+2802 (belonging to the literature sample), whose extreme brightness allows to reach a high SNR also in single frames. We notice that there is a reasonable correlation between the DIMM values and the FWHM measured on the frames. The correlation is stronger with the FWHM of the spectral order spatial profile and, in general, for the higher SNR frames. However, it is also clear that the seeing measured by the DIMM  is too optimistic, being always smaller than the one at the telescope, traced by the FWHM of the order spatial profiles. 
In Fig.~\ref{fig:resol_NIR_J0100} and \ref{fig:resol_NIR_J0439}, we show the same relations but for the NIR frames of SDSSJ0100+2802 (slit=0.6") and J0439+1634 (slit=0.9"), respectively. When considering the relation between the DIMM FWHM and the FWHM of the telluric model (measured in \kms), we converted the former into a velocity width assuming the nominal resolving power associated with the adopted slit, $R$, and a linear relation between the width of the slit and the resolving power: FWHM (\kms)~$= [c /  (R * slit)]$~FWHM ("). For values of the DIMM FWHM larger than the adopted slit, we assumed the nominal resolution ($R_{\rm VIS} \simeq 8900$, $R_{\rm NIR} \simeq 8100$). 

It is interesting to observe how the relation between the FWHM of the telluric model and the DIMM FWHM flattens above a given value (which depends on the XSHOOTER arm and slit width) which however is always smaller than the nominal resolution, suggesting that the latter could be overestimated. 

In order to assign a reasonable resolution to our frames we proceeded through the following steps. 

\begin{enumerate} 
\item The frames with SNR~$\ge 8$ per 10 \kms\ bin are used to build relations between the FWHM of the spatial profiles and the telluric FWHM for each arm and each slit. Data are fitted with a broken line, where the break occurs at the slit width (see Fig.~\ref{fig:resol_all}). 
There is not enough data to fit a relationship for the NIR 1.2" slit, so in this case we adopt the relationship measured from the 0.9" slit. There are only 8 exposures observed with the NIR 1.2" slit, spread across two objects which both have a larger number of exposures at smaller slit widths, so the choice of scaling for this particular configuration is unlikely to have a significant impact on the final results;
\item These relations are used to calculate the expected spectral FWHM for each exposure, based on its measured spatial FWHM;
\item The spectral FWHM of the composite spectrum (obtained summing all the frames of a given object in a given arm) is determined by creating a synthetic composite profile by averaging together the Gaussian profiles obtained from the single spectral FWHM. The individual spectra are inverse-variance weighted when they are combined, so the profiles should be weighted by 1/SNR$^2$;
\item The weighted composite profiles are then fitted with a single Gaussian profile and its FWHM is assumed as the resolution of the combined spectrum. We verified that the composite profiles are very well approximated by single Gaussians. 
\end{enumerate} 

This procedure assumes that the telluric FWHM is a good measure of the spectral resolution for high-SNR exposures, and that the spatial FWHM is more robust than the telluric FWHM for low-SNR exposures. The values obtained for the VIS and NIR regions of the E-XQR-30 sample are reported in Table~\ref{tab:obs_details} and shown in Fig.~\ref{fig:resol_vs_resol}.  Indicatively, the adopted resolution can be considered reliable to within $\approx 10$ \%.

\subsection{Estimate of the intrinsic emission spectrum} 
\label{sec:continuum}

For many science cases it is necessary to determine the shape of the intrinsic spectrum of the quasar. However, this operation is carried out in different ways for different purposes. Here we briefly describe the procedures adopted for the science cases studied with the E-XQR-30 sample referring the reader to the original papers for further details. 

\vspace{0.1cm}
\noindent 
{\it Determination of the transmission flux in the Lyman forests} \\
The emission spectra used to measure the transmission flux and the effective optical depth in the \Lya\ and \Lyb\ forests have been reconstructed employing a Principal Component Analysis (PCA). PCA models use a training set of low-$z$ quasar spectra to find optimal linear decompositions of the ‘known’ red side ($\lambda > 1280$ \AA\ rest frame) and the ‘unknown’ blue side of the spectrum ($\lambda < 1220$ \AA\ rest frame), then determines an optimal mapping between the linear coefficients of the two sides’ decompositions \citep{francis1992,yip2004,suzuki2005,mcdonald2005,durovcikova2020}. 
\citet{bosman2021} showed that PCA methods outperformed both the more traditionally employed power-law extrapolation \citep[e.g.][]{bosman2018} and ‘stacking of neighbours’ methods \citep[e.g.][]{bosmanbecker2015}, both in prediction accuracy and in lack of wavelength-dependent reconstruction residuals. 

For the E-XQR-30 sample, we used the log-PCA approach of \citet{davies2018},  
trained on $\sim4600$ quasars at $2.7 < z < 3.5$ with SNR~$ > 7$ from the SDSS-III Baryon Oscillation Spectroscopic Survey \citep[BOSS,][]{BOSS2013} and the SDSS-IV Extended BOSS \citep[eBOSS,][]{EBOSS2016} and tested using an independent set of $\sim4600$ quasars from eBOSS.

The average uncertainty over the $1026 < \lambda < 1185$ \AA\ range, used in \citet{bosman2022}, was determined from the test set and is PCA/True~$-7.9/+7.8$\%, implying a large improvement compared to power-law extrapolation methods ($> 13$\%) and a slight improvement over the best PCA in \citet{bosman2021} (9\%). 
We refer the reader to \citet{bosman2021} and \citet{bosman2022} for further details of the PCA training and testing procedures. Figures showing all available PCA fits and blue-side predictions are available in \citet{zhu2021}.

\vspace{0.1cm}
\noindent
{\it Fit of intervening metal absorption lines} \\
The emission spectrum for the fit of the metal absorption lines in the region redward of the \Lya\ emission line has been determined with the \textsc{Astrocook} python package \citep{cupani2020}. \textsc{Astrocook} performs first the identification of absorption lines, then  masks the identified lines, calculates the median wavelength and flux in windows of fixed velocity width, and inserts nodes at these locations. Finally, univariate spline interpolation is applied to the nodes to estimate the shape of the continuum at the wavelength sampling of the original spectra. 
The initial continuum estimates are manually refined by
adding and removing nodes as necessary using the \textsc{Astrocook} GUI. This operation is generally needed in regions affected by strong skyline or telluric residuals, when there are large absorption troughs due to clustered narrow absorption lines or to BAL systems or when the spectrum has significantly slope changes in narrow wavelength intervals (e.g. close to emission lines). 
A detailed description of the procedure can be found in \citet{rdavies2023a}. 

\vspace{0.1cm}
\noindent
{\it Identification of BAL quasars and fit of intrinsic emission lines} \\
The intrinsic spectrum of the quasars for the study of the properties related with the quasars themselves (e.g. BH masses, AGN winds) have been estimated with different techniques. 
\citet{bischetti2022}, who determined the BAL fraction in XQR-30,  built composite emission templates based on the catalogue of 11,800 quasars at $2.13 \leq z \leq 3.20$ from SDSS DR7 \citep{shen2011}.  For each XQR-30 spectrum, the template is the median of 100 non-BAL quasar spectra, randomly selected from the SDSS catalogue, with colours and \CIV\  equivalent width in a range encompassing $\pm 20$ \% the values measured for the XQR-30 quasar. The combined template has a spectral resolution of $\approx70$ \kms\ in the \CIV\ spectral region, which corresponds to the lower spectral resolution of the individual SDSS spectra, and is normalized to the median value of the XQR-30 quasar spectrum in the $1650-1750$ \AA\ interval, which is free from prominent emission lines and strong telluric absorption in the X-shooter spectra.

In \citet{lai2022}, on the other hand, in order to detect the chemical abundances in the quasar broad line region (BLR), the continuum is fit with a power-law function normalized to rest-frame 3000 \AA, considering the line-free windows 1445-1455 \AA, 1973-1983 \AA.   We also consider the contribution of the \FeII\ pseudo-continuum spectrum using the empirical template from \citet{vestergaard2001} to cover the wavelength range from 1200 to 3500 \AA. 
Emission lines are fit with a double power-law method \citep{nagao2006} which has been compared to the double-Gaussian and modified Lorentzian methods, achieving better fits with fewer or equal number of free parameters. 

\subsection{Rest-frame UV magnitudes at 1450 \AA}
The quasars' UV monochromatic AB magnitudes at rest-frame 1450 \AA\ ($m_{1450}$) are widely used in the literature. For example, they are used to determine the quasar luminosity function \citep[e.g.][]{schindler2023}, to estimate bolometric luminosities \citep[e.g.][]{runnoe2012}, and to normalise quasar's proximity zones \citep[e.g.][]{eilers2020}.  
At $z > 5.7$, the rest-frame 1450 \AA\ wavelength is shifted to $\lambda>0.97\,\mu$m, so it is not always possible to measure it directly from the observed optical spectrum. Different approaches have been adopted to determine $m_{1450}$, which generally imply extrapolation from the observed spectrum with a power-law or from NIR photometry \citep[see e.g.][]{banados2016}. These determinations are sensitive to the assumed spectral power law and to strong emission/absorption lines affecting the NIR photometry. 


In the case of the E-XQR-30 high-quality spectroscopic sample, we have the possibility to directly measure the flux density at $\lambda_{1450} = 1450 \times (1+z_{\rm em})$ on the flux calibrated spectra and determine $m_{1450}$ without the need of extrapolation. 
To this aim, the flux density, $f_{\lambda 1450}$ (and subsequently, $m_{1450}$) was determined by taking the average value over a few tens of \AA\ of the fit of the continuum, modelled as a power law plus a Balmer pseudo-continuum \citep[following recent works, e.g.][]{schindler2020}. More details on the fit will be described in a subsequent paper (Bischetti et al., in prep.).





\noindent
All $m_{1450}$ magnitudes are reported in Table~\ref{tab:xqr30_sample}. Compared with previous determinations from the literature \citep[see e.g. the high-$z$ database in][]{fan2022}, our magnitudes are fainter in most of the cases with a mean and median discrepancy of $\sim0.2$ magnitudes. This result is reasonable since most previous determinations were based on extrapolations from photometric values (e.g., $y$ or $J$ magnitudes) which are often contaminated by strong emission lines (e.g., \CIV). 


\section{Released data products}
\label{sec:release}

\subsection{Reduced spectra}
All the E-XQR-30 raw data, along with calibration files are available through the ESO archive\footnote{https://archive.eso.org/eso/eso\_archive\_main.html}. 

The spectra reduced with the custom pipeline are released through a public github repository\footnote{https://github.com/XQR-30/Spectra}. For each quasar in the sample, two binary FITS files are released: one with the combined VIS and one with the combined NIR frames. 
The naming convention is {\tt{target\_arm.fits}}, where {\tt{target}} is the target name as reported in Table~\ref{tab:xqr30_sample}, {\tt{arm}} is the spectral arm (NIR or VIS). The table columns are:  
\begin{itemize}
\item[-] WAVE: wavelength in the vacuum-heliocentric system (\AA); 
\item[-] FLUX: flux density with the telluric features removed (erg cm$^{-2}$ s$^{-1}$ \AA$^{-1}$);
\item[-] ERROR: error of the flux density (erg cm$^{-2}$ s$^{-1}$ \AA$^{-1}$); 
\item[-] FLUX\_NOCORR: same as flux, but with the telluric features (erg cm$^{-2}$ s$^{-1}$ \AA$^{-1}$);
\item[-] ERROR\_NOCORR: error of flux\_nocorr (erg cm$^{-2}$ s$^{-1}$ \AA$^{-1}$). 
\end{itemize}

A third data product is released for each quasar which is the combined VIS + NIR spectrum rebinned to 50 \kms\ and normalized to the $J_{\rm AB}$ magnitude. 

\begin{figure*}
\includegraphics[width=\textwidth]{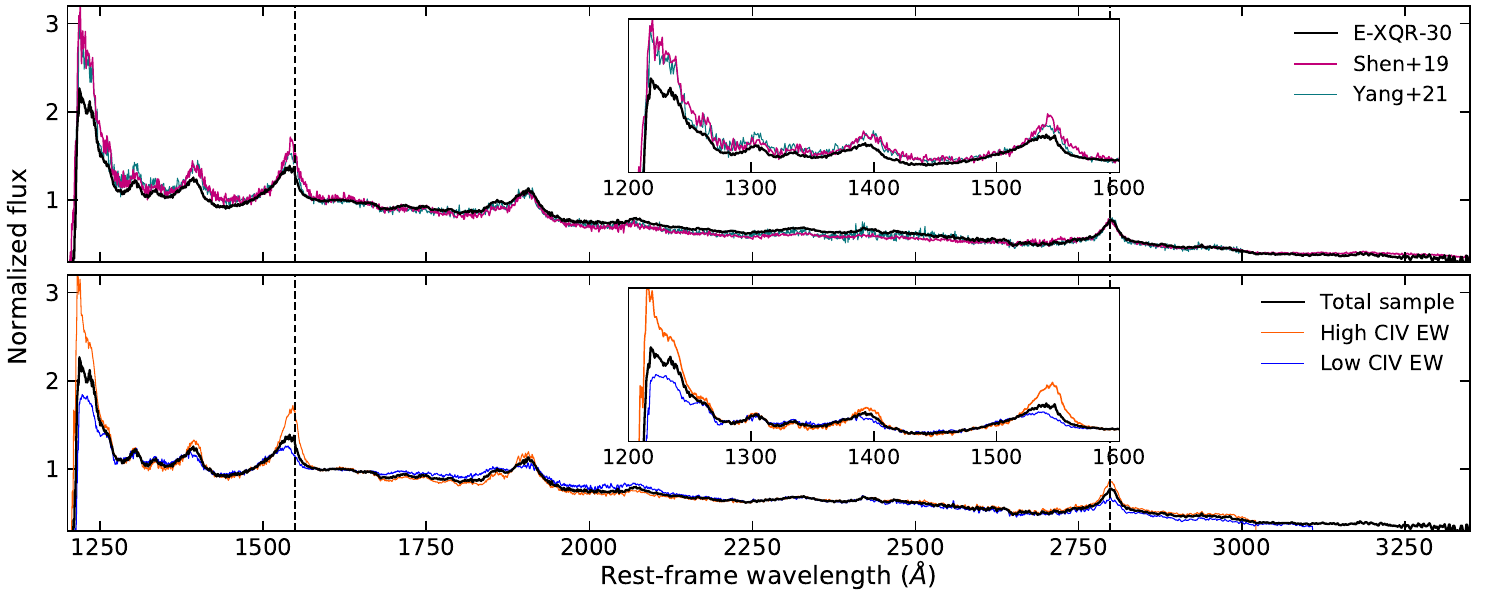}
\caption{\textit{Upper panel:} composite spectrum of the total E-XQR-30 sample rebinned to 250 \kms\ (black curve), compared with composite spectra of $z\simeq5.7-7$ quasars from \citet{shen2019} and \citet{yang2021}. \textit{Bottom panel:} composite E-XQR-30 spectrum for quasars with high (low) equivalent width of the \CIV\ line, as shown by the orange (blue) curve, compared with the total E-XQR-30 composite spectrum. Vertical lines indicate the rest-frame wavelength of \CIV\ and \MgII\ emission lines considering $z_{\rm MgII}$ (Table \ref{tab:xqr30_sample}). Insets show a zoom-in on the spectral region between \Lya\ and \CIV. }
\label{fig:composite}
\end{figure*}

\begin{figure}
\includegraphics[width =8cm]{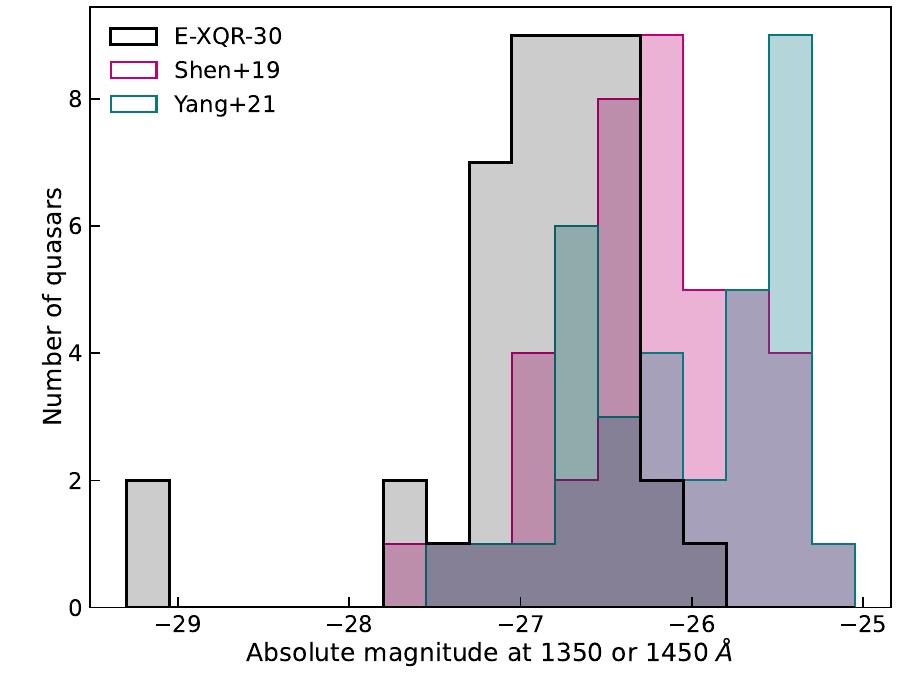}
\caption{Distribution of the absolute magnitude at 1450 \AA\ of the E-XQR-30 sample (black histogram), compared with the quasar samples by \citet{shen2019} and \citet{yang2021}.}
\label{fig:magqso}
\end{figure}

\begin{table}
    \caption{Median composite spectrum of the total E-XQR-30 sample. Wavelengths are in units of \AA\ and flux density units are arbitrary, normalised to the rest-frame 1600-1610 \AA\ continuum flux. Last column indicates the number of quasars contributing to each spectral channel. The entire table data is available online.}
    \centering
    \begin{tabular}{lcc}
    \hline
      Wavelength (\AA)   & $f_\lambda$ & N\\
    \hline
      1150.160 & 0.0015 & 42 \\
      1150.352 & 0.0093 & 42 \\
      1150.543 & 0.0075 & 42 \\
      ...      &     & \\
      1600.468 & 0.9916 & 42 \\
      ...      &     & \\
      3000.856 & 0.4321 & 40\\
      ...      & .    & \\
      3250.427 & 0.3315 & 17 \\
      ...      & ...    & \\
      \hline
    \end{tabular}

    \label{tab:composite}
\end{table}

\subsection{Composite spectra}
\label{sec:composite}

The composite spectrum based on the E-XQR-30 sample was computed to investigate the average quasar UV spectral properties in the redshift range $z\simeq5.8-6.6$ and compare them with previous $z\sim6$ quasar samples from the literature.
We generated a median composite spectrum by converting the individual VIS + NIR combined spectra to the rest-frame using $z_{\rm MgII}$ (Table \ref{tab:xqr30_sample}) and normalizing them to continuum flux at 1600-1610 \AA\ rest frame, as this spectral region is free from prominent emission lines and strong telluric contamination \citep[see][]{shen2019,yang2021}. We did not adopt a normalization at shorter wavelengths \citep[e.g.][]{banados2016} as the spectral region blueward of \CIV\ is affected by absorption in the case of BAL quasars. 

Figure \ref{fig:composite} shows the composite spectrum of the total E-XQR-30 spectrum, corresponding to the spectral region between \Lya\ and \MgII\ emission lines. Shallow absorption is observed blueward of \CIV, due to the large fraction of BAL quasars in our sample \citep{bischetti2022}.  This absorption feature disappears when excluding BAL quasars from the composite (see Fig.~\ref{fig:comp_noBAL_nopDLA}). We do not find other significant differences between the composite spectra including or excluding BAL quasars. 
We also verified that no significant discrepancies arise when comparing the total composite with that created excluding the quasar spectra showing a strong \HI\ absorber within $\simeq 5000$ \kms\ from the emission redshift (a so-called proximate Damped \Lya\ system, see \citealt{rdavies2023a} and Sodini et al. in prep.). The comparison is shown in Fig.~\ref{fig:comp_noBAL_nopDLA}.

The composite E-XQR-30 spectrum covers the spectral region from rest-frame 1150 to 3350 \AA. We did not include shorter wavelengths, due to the strong IGM absorption. The total sample of 42 quasars contribute to the composite at wavelengths below 3000 \AA\ while the number of quasars included in the composite decreases at longer wavelengths. The composite has a relatively low quality at $>3270$ \AA, where less than 10 quasars contribute to the obtained flux.

We compare the composite E-XQR-30 spectrum with the median composites from \cite{shen2019} and \cite{yang2021}. The former is based on 50 quasars at $z\simeq5.7-6.4$ with Gemini/GNIRS spectroscopy ($R\sim650$); the latter consists of 32 quasars at $z\simeq6.5-7.6$, the majority of which was observed with GNIRS and Keck/NIRES ($R\sim2700$). Both samples include bright quasars with similar magnitudes to E-XQR-30, with a few common sources, but mostly consist  of less luminous quasars (Fig.~\ref{fig:magqso}).

The composite spectrum of E-XQR-30 shows less prominent \Lya\ and high-ionization emission lines such as \CIV\ and \SiIV\ with respect to \cite{shen2019} and \cite{yang2021} composites, while there is no significant difference in the strength of \MgII\ and \CIII]. In our composite, the peak of the \CIV\ emission shows a larger blueshift and the line profile is characterized by a more prominent blue wing, tracing the presence of strong nuclear winds in the quasar BLR, whose properties will be characterized in a forthcoming paper (Bischetti et al. in prep). This is likely due to the luminosity difference between E-XQR-30 and the \cite{shen2019,yang2021} samples \citep[e.g.][]{vietri2018,schindler2020}. The three composites have consistent continuum slopes. 

To highlight the variety of quasar UV properties within our sample, we also generated two additional composites for sources with low ($<14.5$ \AA) and high ($>14.5$ \AA) \CIV\ equivalent width (EW). Sources with low (high) \CIV\  EW  typically have weaker (stronger) \Lya, \SiIV\ and \MgII\ emission lines. The interpretation of differences close to \CIII] is limited due to strong telluric contamination affecting this spectral region. Quasars with low \CIV\ EW typically show \CIV\ blueshifts with velocities $\geq2000$ \kms\ (Bischetti et al. in prep.), consistently with the \CIV\ EW-\CIV\ blueshift anticorrelation typically found in lower-redshift quasars \citep{richards2011,temple2021}.

The three composite spectra shown in Fig.~\ref{fig:composite} are part of the released products, they can be found in the github repository where also the reduced spectra are stored. An example of the format of the released tables is shown in  Table~\ref{tab:composite}. 

\subsection{Metal absorption line catalogue}
\label{sec:catalogue}

\citet{rdavies2023a} have searched the E-XQR-30 quasar spectra for metal absorption lines in the region redward of the \Lya\ emission line of the quasar, compiling a catalogue of 778 systems spanning the redshift range $2 \lsim z \lsim 6.5$. Each system shows absorption in one or more of the following ions: \MgII\ (360 systems), \FeII\ (184), \CII\ (46), \CIV\ (479), \SiIV\ (127), and \NV\ (13). This catalog significantly expands on existing samples of $z \gsim 5$ absorbers, especially for \CIV\ and \SiIV\ which are important probes of the ionizing photon background at high redshift. The catalogue has been released through the public XQR-30 github repository\footnote{https://github.com/XQR-30/Metal-catalogue} along with completeness statistics and a Python script to compute the absorption search path for different ions and redshift ranges.

\section{Summary of published results}
\label{sec:result}

At the time of writing, several results  have been published by the XQR-30 collaboration based on large quasar samples mainly built on the high-quality E-XQR-30 spectra. We briefly summarize them in the following. 

The last phases of the reionization process have been studied with several statistical indicators derived from the quasar spectra. The main result, common to all studies, is that when comparing observables with simulations a fully ionized IGM with a homogeneous UVB is disfavored by the data down to $z \simeq 5.3$.
In \citet{bosman2022}, an improved measurements of the mean \Lya\ transmission in the redshift range $4.9 < z < 6.1$ is derived from a sample of 67 quasar sightlines with $z_{\rm em} > 5.5$. The sample includes the E-XQR-30 spectra (excluding 5 objects with strong BAL systems and QSO J0439+1634 whose XSHOOTER spectrum was not available at the time) plus other 15 XSHOOTER and 16 Keck ESI quasar spectra. We find excellent agreement between the observed \Lya\ transmission distributions and the homogeneous-UVB simulations Sherwood \citep{bolton2017} and Nyx \citep{almgren2013} up to $z \leq 5.2$ ($< 1\, \sigma$), and mild tension ($\sim2.5\, \sigma$) at $z=5.3$. Homogeneous UVB models are ruled out by excess \Lya\ transmission scatter at $z \ge 5.4$ with high confidence ($> 3.5\, \sigma$). Our results indicate that reionization-related fluctuations, whether in the UVB, residual neutral hydrogen fraction, and/or IGM temperature, persist in the IGM until at least $z = 5.3$ ($t = 1.1$ Gyr after the Big Bang), strongly suggesting a late end to reionization.

Two works investigate the statistics of the "dark gaps", in the \Lya\ and \Lyb\ forests \citep[][respectively]{zhu2021,zhu2022}.  Dark gaps are contiguous regions of strong absorption which could be created by regions of neutral IGM and/or low UV background \citep[e.g.][]{nasir2020}.
For the analysis in the \Lya\ forest, we use high-SNR spectra of 55 quasars at $z_{\rm em} > 5.5$  taken with XSHOOTER (35 belonging to E-XQR-30) and Keck ESI. Focusing on the fraction of sightlines containing dark gaps of length $L\ge 30\,h^{-1}$ Mpc as a function of redshift, $F_{30}$, we measure $F_{30} \simeq 0.9, 0.6$, and 0.15 at $z = 6.0, 5.8$, and 5.6, respectively, with the last of these long dark gaps persisting down to $z \simeq 5.3$. Furthermore, nine ultralong ($L> 80 h^{-1}$ Mpc) dark gaps are identified at $z < 6$. The presence of long dark gaps at these redshifts demonstrates that large regions of the IGM remain opaque to \Lya\ down to $z \simeq5.3$.
For what concern the \Lyb\ forest, the sample is reduced to the 42 quasars with the best SNR at $5.77 \leq z_{\rm em} \leq 6.31$.  We show that about 80\,\%, 40\,\%, and 10\,\% of quasar spectra exhibit long ($L \ge 10 h^{-1}$ Mpc) dark gaps in their \Lyb\ forest\footnote{Note that \Lyb\ dark gaps are required to be dark in the \Lya\ forest too.} at $z \simeq 6.0, 5.8$, and 5.6, respectively. Among these gaps, we detect a very long ($L= 28 h^{-1}$ Mpc) and dark ($\tau_{\rm eff} \gsim 6$) \Lyb\ gap extending down to $z \simeq5.5$ toward the $z_{\rm em} = 5.816$ quasar PSO~J025-11. 
Finally, we infer constraints on $\langle x_{\rm HI} \rangle$ over $5.5 \lsim z \lsim 6.0$ based on the observed \Lyb\ dark gap length distribution and a conservative relationship between gap length and neutral fraction derived from simulations. We find  $\langle x_{\rm HI} \rangle \le 0.05, 0.17$, and 0.29 at $z \simeq 5.55, 5.75$, and 5.95, respectively. 

The proximity zone, due to the enhanced radiation near a luminous quasar, is the only region where it is possible to characterize the density field in high-redshift quasar spectra which otherwise show heavily saturated \Lya\ absorption. Using a sample of 10 quasar spectra from the E-XQR-30 survey, \citet{chen2022} measure the density fields in their proximity zones out to $\sim 20$ cMpc for the first time. The recovered density cumulative distribution function (CDF) is in excellent agreement with the modeled one from the CROC simulation \citep{gnedin2014,chen2021} between 1.5 and 3 pMpc from the quasar, where the halo-mass bias is low. 
This region is far away from the quasar hosts and hence approaching the mean density of the Universe, which allows us to use the CDF to set constraints on the cosmological parameter $\sigma_8 = 0.6 \pm 0.3$. The uncertainty is mainly due to the limited number of high-quality quasar sightlines currently available. In the region closer to the quasar, within 1.5 pMpc, we find that the density is higher than predicted in the simulation by $1.23 \pm 0.17$, suggesting that the typical host dark matter halo mass of a bright quasar ($M_{1450} <-26.5$) at $z \sim 6$ is $\log_{10} M_h / M_{\sun} = 12.5^{+0.4}_{-0.7}$. 

While detecting individual \HI\ \Lya\ forest lines in spectra of high-redshift quasars is difficult, heavy element absorbers are readily detected redward of the \Lya\ emission. They are fundamental probes of the ionization state and chemical composition of the circumgalactic and intergalactic medium near the end of the EoR. In \citet{rdavies2023a}, we have carried out a systematic analysis in all the quasar spectra of the E-XQR-30 sample to create a catalogue of 778 intervening absorption systems in the range $2 \lsim z \lsim 6.5$ with Voigt fit parameters and completeness estimates. The catalogue of absorbers together with all parameters derived from the Voigt profile fitting have been released to the public (see Sec.~\ref{sec:release}). The redshift evolution of the statistical properties of \CIV\ absorption lines over $4.3 \lsim z \lsim 6.3$ is presented in \citet{rdavies2023b}.  
We find that the \CIV\ cosmic mass density ($\Omega_{\rm CIV}$) decreases by a factor of $4.7 \pm 2.0$ over the $\sim 300$ Myr interval between $z \simeq 4.7$ and $z \simeq 5.8$. Assuming that the carbon content of the absorbing gas evolves as the integral of the cosmic star formation rate density (with some time delay due to stellar lifetimes and outflow travel times), we show that chemical evolution alone could plausibly explain the observed fast decline in  $\Omega_{\rm CIV}$. However, our data also reveal evidence for a decrease in the \CIV/\CII\ ratio at the highest redshifts \citep[see also][]{cooper2019}. Rapid changes in the ionization state of the absorbing gas driven by the evolution of the UV background towards the end of hydrogen reionization \citep[see also][]{becker2019} may contribute to the increased rate of decline in $\Omega_{\rm CIV}$ at $z \gsim 5$.
 
The intrinsic properties of the XQR-30 quasars have been investigated in \citet{bischetti2022}, where we studied the incidence and characteristics of broad absorption line (BAL) systems. BAL are absorption lines characterized by width of thousands of \kms\ and velocities from the systemic emission redshift that can reach $0.1-0.2 c$ \citep[e.g.][]{weymann1983}. They trace strong BH-driven outflows that are believed to have the potential to contribute to AGN feedback  \citep[e.g.][]{fiore2017}. We found that about $40-47$\,\% of XQR-30 quasars show BAL features identified through the \CIV\ absorption, a fraction which is $\sim2.4$ times higher than what we measure in $z \sim 2 - 4$ SDSS quasars selected in a consistent way. 
Furthermore, the majority of BAL outflows at $z \sim 6$ also show extreme velocities of $20,000-50,000$ \kms, rarely observed at lower redshift. These results indicate that BAL outflows in $z \sim 6$ quasars inject large amounts of energy in their surroundings and may represent an important source of BH feedback at these early epochs, possibly driving the transition from a growth phase in which BH growth is dominant to a phase of BH and host-galaxy symbiotic growth \citep{volonteri2012}. 

In \citet{lai2022}, the intrinsic emission lines in the spectra of 25 high-redshift ($z_{\rm em} >5.8$) quasars (15 from XQR-30) are used to estimate the \CIV\ blueshifts and the BLR metallicity-sensitive line ratios.  
Comparing against CLOUDY-based photoionization models, the metallicity inferred from line ratios in this study is several (at least $2-4$) times supersolar,  consistent with studies of much larger samples at lower redshifts and similar studies at comparable redshifts. We also find no strong evidence of redshift evolution in the BLR metallicity, indicating that the BLR is already highly enriched at $z \sim 6$. Our high-redshift measurements also confirm that the blueshift of the \CIV\ emission line is correlated with its equivalent width, which influences line ratios normalized against \CIV. When accounting for the \CIV\ blueshift, we find that the rest-frame UV emission-line flux ratios do not correlate appreciably with the BH mass or bolometric luminosity. 

In \citet{satyavolu2023b}, we have measured proximity zone sizes of 22 quasars from the E-XQR-30 sample. In our analysis, we exclude quasars with BALs  and proximate DLAs that could contaminate or truncate the proximity zone size and lead to a spurious measurement. We infer proximity zone sizes between $2-7$ physical Mpc, with a typical error of less than 0.5~physical Mpc, which includes for the first time, the errors due to the uncertainty in the quasar continuum.  We compare the distribution of our proximity zone sizes to those from simulations \citep{satyavolu2023}; study the correlation between proximity zone sizes and the quasar redshift, luminosity, or black hole mass, and find that they indicate a large diversity of quasar lifetimes. 
We find two quasars with exceptionally small proximity zone sizes ($< 2$~physical Mpc). The spectrum of one of these quasars,  PSOJ158-14, at $z_{\rm em} = 6.0685$, shows, unusually for this redshift, damping wing absorption without any proximate metal absorbers, which could potentially originate from the IGM. The other quasar, PSOJ108+08, has a high-ionization absorber at $\sim$0.5~physical~Mpc from the edge of the proximity zone. 

The BH masses and accretion rates, and the bolometric luminosities for all the objects in the E-XQR-30 sample are computed in Mazzucchelli et al. (subm.). Two estimates of the BH mass were derived for each object based on the measured FWHM of \CIV\ and \MgII\ emission lines.   
In this work, we report only the masses derived from \MgII\ (see Table~\ref{tab:xqr30_sample}) which are generally considered to be more reliable due to the out-flowing components and/or winds which commonly affects the \CIV\ emission line profile.

\section{Conclusions}
\label{sec:conclusion}

In this paper, we present the XQR-30 survey: a sample of 30 spectra of quasars probing the reionization epoch ($5.8 \leq z_{\rm em} \leq 6.6$) observed at high resolution ($R\sim 10,000$) and high signal-to-noise ratio (SNR~$\sim 11-41$ per bin of 10 \kms)   within an ESO XSHOOTER Large Programme of 248 h. 
The XQR-30 observations were complemented with all the available XSHOOTER archival observations for the objects in the sample and with the addition of 12 more quasars with archival XSHOOTER spectra of similar quality, for a total of $\sim350$ hours on-source observation. The total sample of 42 quasars is designated as the ``enlarged XQR-30" sample  (E-XQR-30). 

 E-XQR-30  represents the state-of-the-art observational database for the studies  of the second-half of the reionization process with quasar spectra.

Many results have been already published based on this sample, spanning from the statistics of the transmitted flux in the \Lya\ and \Lyb\ forest strongly constraining reionization models \citep{bosman2022,zhu2021,zhu2022}; the reconstruction of the high-$z$ density field in the quasar proximity regions \citep{chen2022}; the detection, identification and analysis of heavy element absorption lines  \citep{rdavies2023a,rdavies2023b}; and the intrinsic properties of the luminous quasars in the sample being powered by super-massive BH and showing the signatures of strong outflows possibly driving the transition to the co-evolution of the BH and host-galaxy masses \citep[][Mazzucchelli et al. 2023 subm.]{lai2022,bischetti2022,bischetti2023,satyavolu2023b}. 
Several other papers are in preparation and will be submitted soon. 

This sample is intended to have a high community value: the reduced spectra are made available through a public repository, together with the composite spectra and the catalogue of metal absorption lines. Several relevant properties of the observed quasars are reported in this and in the other published papers. 
Furthermore, XQR-30 has a notable legacy relevance since it is improbable that such an investment of telescope time will be repeated on similar targets in the next 5-10 years. Given the finite number of bright quasars with $z>6$ in the observable Universe \citep{fan2022}, we may have to wait for the next generation of 30-40 m class telescopes equipped with high-resolution spectrographs  \citep[e.g. the ANDES spectrograph for the ESO ELT,][]{marconi2021} to mark a significant step forward in this field. 
 
\section*{Acknowledgements}
We are grateful to the referee, Prof. Michael Strauss, for his comments and suggestions which have allowed us to correct some mistakes and improve the clarity of the paper.    
SEIB, RAM and FW acknowledges funding from the European Research Council (ERC) under the European Union’s Horizon 2020 research and innovation programme (Grant agreement No. 740246 “Cosmic Gas”).
HC is supported by the Natural Sciences and Engineering Research Council of Canada (NSERC), funding reference \#DIS-2022-568580. 
Parts of this research were supported by the Australian Research Council Centre of Excellence for All Sky Astrophysics in 3 Dimensions (ASTRO 3D), through project number CE170100013. 
EPF is supported by the international Gemini Observatory, a program of NSF’s NOIRLab, which is managed by the Association of Universities for Research in Astronomy (AURA) under a cooperative agreement with the National Science Foundation, on behalf of the Gemini partnership of Argentina, Brazil, Canada, Chile, the Republic of Korea, and the United States of America.
AF acknowledges support from the ERC Advanced Grant INTERSTELLAR H2020/740120.
Support by ERC Advanced Grant 320596 ‘The Emergence of Structure During the Epoch of reionization’ is gratefully acknowledged by MGH.
SRR acknowledges financial support from the International Max Planck Research School for Astronomy and Cosmic Physics at the University of Heidelberg (IMPRS--HD).
JTS acknowledges funding from the ERC Advanced Grant program under the European Union’s Horizon 2020 research and innovation programme (Grant agreement No. 885301).
RT acknowledges financial support from the University of Trieste. RT acknowledges support from PRIN MIUR project “Black Hole winds and the Baryon Life Cycle of Galaxies: the stone-guest at the galaxy evolution supper”, contract \#2017PH3WAT.
This research was supported in part by the National Science Foundation under Grant No. NSF PHY-1748958. 
For the purpose of open access, the authors have applied a Creative Commons Attribution (CC BY) licence to any Author Accepted Manuscript version arising from this submission.

This paper is based on the following ESO observing programmes: XSHOOTER 60.A-9024, 084.A-0360, 084.A-0390, 084.A-0550, 085.A-0299, 086.A-0162, 086.A-0574, 087.A-0607, 088.A-0897, 091.C-0934, 294.A-5031, 096.A-0095, 096.A-0418, 097.B-1070, 098.B-0537, 0100.A-0625, 0101.B-0272, 0102.A-0154, 0102.A-0478, 1103.A-0817;  HAWKI (H band data) 105.20GY.005.

\section*{Data Availability}

 The reduced spectra, the flux calibrated spectra and the composites described in Section~\ref{sec:release} are released with this paper and can be downloaded from this GitHub repository: https://github.com/XQR-30/.



\bibliographystyle{mnras}
\bibliography{biblio_vale}




\appendix

\section{Resolution estimate} 

\begin{figure*}
	\includegraphics[width=16cm]{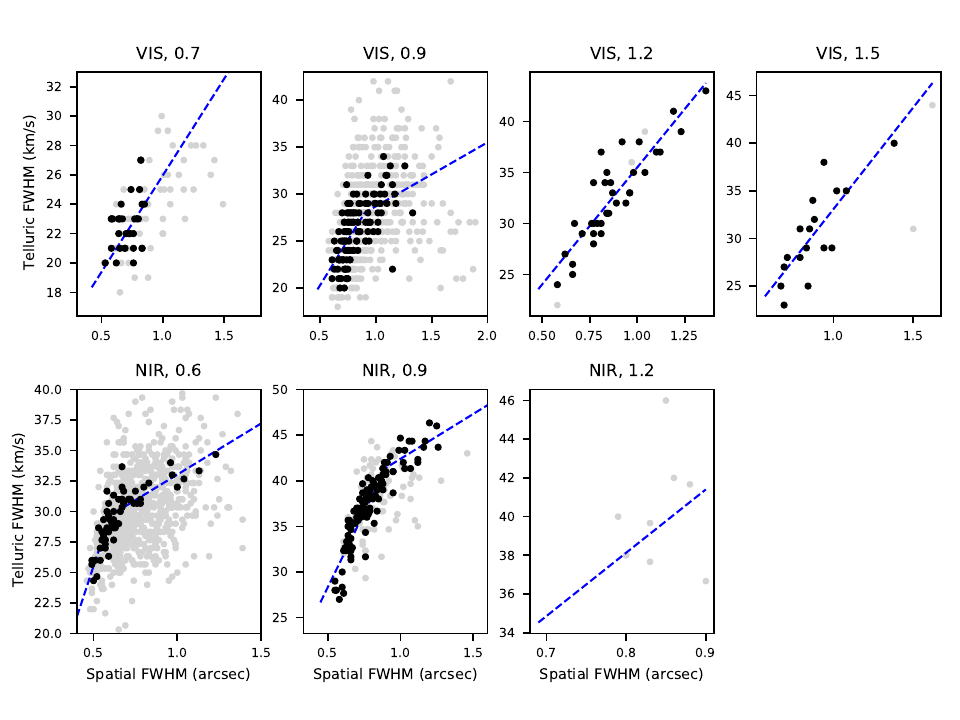}
    \caption{FWHM of the telluric model measured in \kms\ vs. the FWHM of the spectral order spatial profile in arcsecs for the VIS (upper row) and NIR (lower row) XSHOOTER arms and for all the slits adopted in the considered frames. The width of the slit in arcsec is reported in the upper part of every panel. The black dots are the frames with SNR~$\ge 8$ per 10 \kms\ pixel, the grey dots are the frames with lower SNR. The dashed blue line is fitted to the black dots with a break at the slit value. See Section~\ref{sec:resol} for more details.}
    \label{fig:resol_all}
\end{figure*}

\section{Composite spectra}

\begin{table}
	\centering
	\caption{List of the targets not showing BAL systems based on the work by \citet{bischetti2022}, contributing to the no-BAL composite spectrum shown in Fig.~\ref{fig:comp_noBAL_nopDLA} with the orange curve. The reported emission redshift is  measured with [\CII], when available, and with \MgII\ otherwise. Refer to Table~\ref{tab:xqr30_sample} for the other general properties of the quasars.  }
	\label{tab:noBAL}
	\begin{tabular}{ll} 
		\hline
Target name  & $z_{\rm em}$ \\
\hline
SDSS J0927+2001  & 5.7722\\
PSO J308-27    & 5.799 \\
SDSS J0836+0054  & 5.810 \\
PSO J242-12     & 5.840 \\
PSO J025-11    & 5.849    \\
PSO J108+08   & 5.955     \\
ULAS J0148+0600   & 5.977 \\
PSO J029-29    & 5.981 \\
SDSS J0818+1722  & 5.997 \\
PSO J007+04    & 6.0015   \\
VST-ATLAS J029-36  & 6.013 \\
SDSS J1306+0356  & 6.0330  \\
PSO J158-14   & 6.0685 \\
CFHQS J1509-1749 & 6.1225 \\
ULAS J1319+0950  & 6.1347 \\
PSO J217-16    & 6.1498 \\
PSO J060+24    & 6.170   \\
PSO J359-06    & 6.1722  \\
PSO J065-26    & 6.1871    \\
SDSS J1030+0524  & 6.304 \\
SDSS J0100+2802  & 6.3268 \\
VST-ATLAS J025-33 & 6.3373 \\
DELS J1535+1943   & 6.381 \\
PSO J183+05    & 6.4386 \\
WISEA J0439+1634  & 6.5188 \\
VDES J0224-4711   & 6.525 \\
PSO J036+03   & 6.5405  \\
PSO J323+12   &  6.5872 \\
\hline 
	\end{tabular}
\end{table}

\begin{table}
	\centering
	\caption{List of the 36 targets showing no Damped \Lya\ systems (DLAs) within $\simeq 5000$ \kms from the emission redshift of the quasar, used to create the no-pDLA composite shown in Fig.~\ref{fig:comp_noBAL_nopDLA} with the blue curve. DLAs at these redshifts cannot be identified trough the detection of the characteristic \Lya\ absorption line showing the damping wings of the Lorentzian profile, they were selected through the detection of the absorption line due to the \OI\ $\lambda\, 1302$ \AA\ transition \citep[][ Sodini et al. in prep.]{rdavies2023a}. The reported emission redshift is  measured with [\CII], when available, and with \MgII\ otherwise. Refer to Table~\ref{tab:xqr30_sample} for the general properties of the quasars. }
	\label{tab:nopDLA}
	\begin{tabular}{ll} 
		\hline
Target name  & $z_{\rm em}$ \\
\hline
SDSS J0927+2001  & 5.7722\\
PSO J308-27    & 5.799 \\
PSO J065+01    & 5.804    \\
SDSS J0836+0054  & 5.810 \\
PSO J023-02      & 5.817 \\
PSO J242-12     & 5.840 \\
PSO J183-12   & 5.893 \\
PSO J108+08   & 5.955     \\
PSO J089-15   & 5.972    \\
ULAS J0148+0600   & 5.977 \\
PSO J029-29    & 5.981 \\
VDES J2250-5015  & 5.985 \\
SDSS J0818+1722  & 5.997 \\
PSO J009-10    & 6.0040   \\
VST-ATLAS J029-36  & 6.013 \\
SDSS J1306+0356  & 6.0330  \\
VDES J0408-5632  & 6.033  \\
PSO J158-14   & 6.0685 \\
SDSS J0842+1218   & 6.0754 \\
PSO J239-07   & 6.1102 \\
CFHQS J1509-1749 & 6.1225 \\
PSO J217-16    & 6.1498 \\
PSO J217-07    & 6.166 \\
PSO J060+24    & 6.170   \\
PSO J359-06    & 6.1722  \\
SDSS J1030+0524  & 6.304 \\
SDSS J0100+2802  & 6.3268 \\
VST-ATLAS J025-33 & 6.3373 \\
VDES J2211-3206   & 6.3394 \\
DELS J1535+1943   & 6.381 \\
WISEA J0439+1634  & 6.5188 \\
VDES J0224-4711   & 6.525 \\
PSO J036+03   & 6.5405  \\
PSO J231-20   & 6.5869 \\
PSO J323+12   &  6.5872 \\
DELS J0923+0402 &   6.6330 \\ 
\hline 
	\end{tabular}
\end{table}

\begin{figure*}
	\includegraphics[width=16cm]{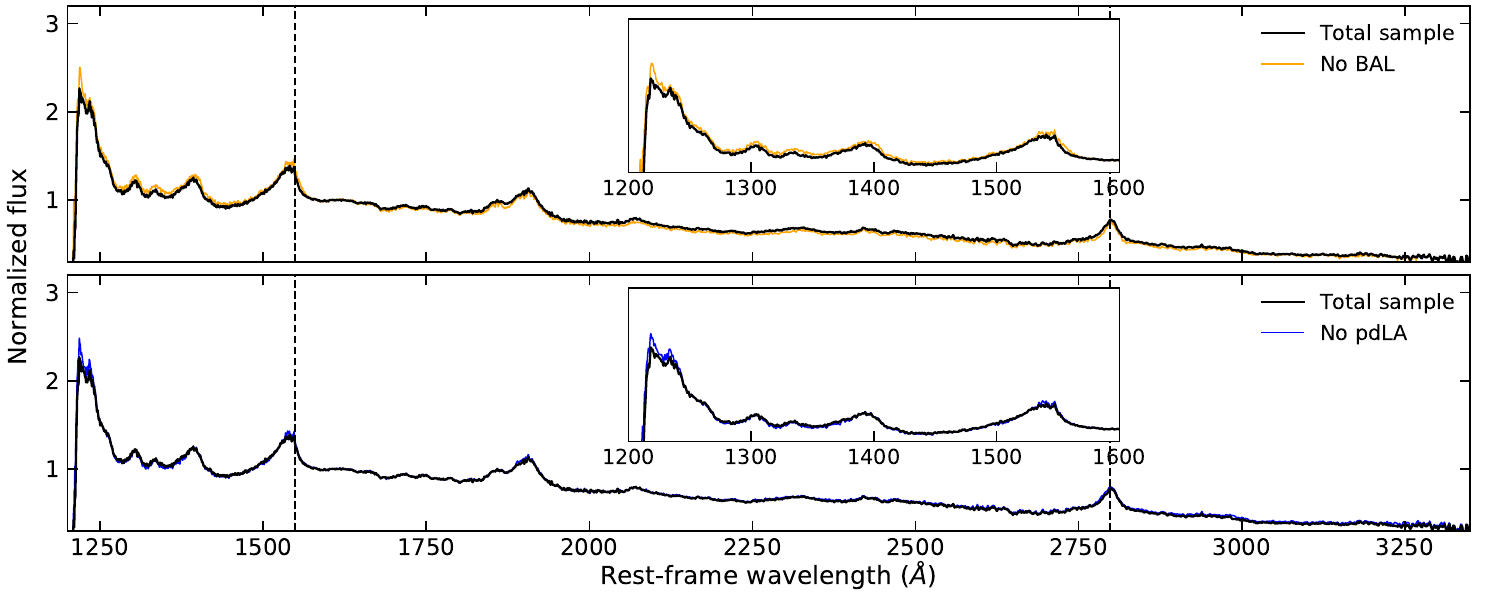}
    \caption{\textit{Upper panel:} Composite spectrum of the total E-XQR-30 sample rebinned to 250 \kms\ (black curve), compared with the composite spectrum obtained excluding the targets showing BAL systems (orange curve), as detected in \citet{bischetti2022}. The list of targets contributing to the no-BAL composite is reported in Table~\ref{tab:noBAL}. \textit{Lower panel:} Composite spectrum of the total E-XQR-30 sample rebinned to 250 \kms\ (black curve), compared with the composite spectrum obtained excluding the targets showing proximate DLAs (blue curve), as identified in \citet{rdavies2023a} and Sodini et al. in prep. The list of targets contributing to the no-pDLA composite is reported in Table~\ref{tab:nopDLA}.}
    \label{fig:comp_noBAL_nopDLA}
\end{figure*}



\bsp	
\label{lastpage}
\end{document}